\documentclass[11pt,letterpaper]{article}
\usepackage{mathtools}
\usepackage{axodraw2}
\usepackage{tikz}
\usepackage{tikz-feynman}
\usepackage{amscd}
\usepackage{amsmath}
\usepackage{slashed}
\usepackage{gensymb}
\usepackage{bm}
\usepackage{jheppub} 

\newcommand{\BthreeLtwo}{${B_3-L_2}$ $Z^\prime$ model}
\newcommand{\bsll}{$b \rightarrow s \mu^+ \mu^-$}
\newcommand{\ZP}{Z^\prime}
\newcommand{\smelli}{{\tt smelli2.3.2}}
\newcommand{\flavio}{{\tt flavio2.3.3}}
\newcommand{\C}{\mathbb{C}}

\newcommand\BthreeLtwogzp{{0.239}}
\newcommand\BthreeLtwotheta{{-0.0133}}
\newcommand\BthreeLtwodchisq{{3.6}}

\newcommand\Sthreegzp{{1.3}}
\newcommand\Sthreetheta{{-0.000467}}
\newcommand\Sthreedchisq{{3.6}}
\title{\centering The Rumble in the Meson:\\ a leptoquark versus a {\boldmath $Z^\prime$}\\
  to fit {\boldmath $b\rightarrow s \mu^+ \mu^-$} anomalies
\\including 2022 LHCb {\boldmath $R_{K^{(\ast)}}$} measurements}

\preprint{CERN-TH-2022-181}
\author[a,b]{Ben Allanach\note{Corresponding author}}
\affiliation[a]{DAMTP, University of Cambridge, Wilberforce Road, Cambridge, 
  CB3 0WA, United Kingdom}
\affiliation[b]{Department of Theoretical Physics, CERN, 1211 Geneva 23, Switzerland}
\author[c,1]{and Joe Davighi}
\affiliation[c]{Department of Physics, Winterthurerstrasse 190, University of
  Z\"urich, CH-8057 Z\"urich, Switzerland}

\emailAdd{B.C.Allanach@damtp.cam.ac.uk}
\emailAdd{joedavighi@gmail.com}

\abstract{We juxtapose global fits of two bottom-up models (an $S_3$ scalar leptoquark model and a ${B_3-L_2}$ $Z^\prime$ model) of \bsll\ anomalies to flavour data  in order to quantify statistical preference or lack thereof. The leptoquark model couples directly to left-handed di-muon pairs, whereas the $Z^\prime$ model couples to di-muon pairs with a vector-like coupling. $B_s-\overline{B_s}$ mixing is a focus because it is typically expected to disfavour $Z^\prime$ explanations. In two-parameter fits to 247 flavour observables, including $B_{s/d} \to \mu^+ \mu^-$ branching ratios for which we provide an updated combination and LHCb $R_{K^{(\ast)}}$ measurements from December 2022, we show that each model provides a similar improvement in quality-of-fit of $\sqrt{\Delta \chi^2}=3.6$ with respect to the Standard Model. The main effect of the $B_s-\overline{B_s}$ mixing constraint in the $Z^\prime$ model is to disfavour values of the $s_L-b_L$ mixing angle greater than about $5|V_{cb}|$. This limit is rather loose, meaning that a good fit to data does not require `alignment' in either quark Yukawa matrix. No curtailment of the $s_L-b_L$ mixing angle is evident in the $S_3$ model.}  


\begin{document}

\maketitle
\flushbottom

\section{Introduction \label{sec:intro}}

Tensions between Standard Model (SM) predictions and measurements of some neutral current
flavour-changing $B-$meson decays persist. We collectively call these tensions
the \bsll\ anomalies.
For example, various lepton flavour universality (LFU) observables like the
ratios of branching ratios
$R_{K^{(\ast)}}=BR(B \rightarrow K^{(\ast)} \mu^+ \mu^-)/BR(B \rightarrow
K^{(\ast)} e^+ e^-)$
were, prior to December 2022, measured to be lower than their SM predictions in multiple channels and
several bins of di-lepton invariant mass
squared~\cite{LHCb:2017avl,LHCb:2019hip,LHCb:2021trn}.
Measurements of other
similar
ratios in $B^0 \rightarrow K^0_s \ell^+ \ell^-$ and $B^\pm \rightarrow
K^{\ast\pm} \ell^+ \ell^-$ decays~\cite{LHCb:2021lvy} (where $\ell \in \{ e,
\mu \}$) are compatible with a
concomitant deficit in the di-muon channel as compared to the di-electron
channel, although we note that the statistics are diminished in the $K_s^0$ and
${K^\ast}^{\pm}$ channels and the tension is not significant in them (it is
around the 1$\sigma$ level only).
SM predictions of all of the double ratios mentioned above enjoy rather
small theoretical  
uncertainties due to their cancellation between numerator and denominator in each case for
the di-lepton invariant mass squared bins of interest,
 such that they are robust.  
We include all of the aforementioned observables in the `LFU' category,
following the computer program\footnote{We use the development version of
  \smelli{} that was on {\tt github} on 27/4/22. We have then updated the
  experimental constraints coming from recent CMS $\overline{BR}(B_s\rightarrow
  \mu^+\mu^-)$ and $BR(B\rightarrow \mu^+\mu^-)$
  measurements~\cite{CMS-PAS-BPH-21-006} and re-calculated the covariances as discussed in 
  \S\ref{sec:bsmm}.} {\tt
  smelli2.3.2}~\cite{Aebischer:2018iyb} that we 
shall 
use to calculate them.
In December 2022, LHCb released a reanalysis of the $R_{K^{(\ast)}}$
  measurements including experimental systematic effects that were missing
  from the previous analysis and a tighter selection of electrons. LHCb has
  decreed that the reanalysis supplants previous results.
  The reanalysis implies
  that the
  four measurements of $R_{K^{(\ast)}}$ are actually \emph{compatible} with SM
  predictions~\cite{LHCb:2022qnv}, contrary to LHCb's previous analyses.

After much theoretical
work, $\overline{BR}(B_s \rightarrow \mu^+ \mu^-)$\footnote{The bar in $\overline{BR}(B_s \rightarrow \mu^+ \mu^-)$ denotes
  an average over the $CP$-untagged, time
  integrated decays~\cite{Alonso:2014csa}.}
also has quite small
theoretical uncertainties in its SM prediction but, together with the
correlated measurement of $BR(B \rightarrow \mu^+ \mu^-)$, it
displays a 1.6$\sigma$ tension
with a combination of the experimental
measurements~\cite{ATLAS:2018cur,CMS-PAS-BPH-21-006,LHCb:2017rmj}.
Several other observables in $B-$meson decays appear to be in significant
tension with SM 
measurements even when one takes into account their larger theoretical
uncertainties, for example $BR(B^+\rightarrow
K^+\mu^+\mu^-)$~\cite{LHCb:2014cxe,Parrott:2022rgu}, $BR(B_s \rightarrow \phi \mu^+
\mu^-)$~\cite{LHCb:2015wdu,CDF:2012qwd} and  angular distributions in $B\rightarrow K^\ast \mu^+ \mu^-$
decays~\cite{LHCb:2013ghj,LHCb:2015svh,ATLAS:2018gqc,CMS:2017rzx,CMS:2015bcy,Bobeth:2017vxj}. For
these quantities,
which are put in the `quarks' category of observable, there 
is room for argument about the best predictions and the
size of the associated theoretical uncertainties~\cite{Gubernari:2022hxn}\footnote{Note that
  discrepancies between predictions and data are still present when one uses
  ratios 
  of observables to cancel their dependence upon CKM matrix elements,
  whose extraction from
  data is based on $\Delta M_{s,d}$, $\epsilon_K$, $S_{\psi K_S}$
  for which we find negligible new physics contributions~\cite{Buras:2022wpw,Buras:2022qip}.}.

\begin{table}\begin{center}
  \begin{tabular}{|c|cccc|}\hline
category & $n_{obs}$ & $\chi^2_{SM}$ & $p$ & $s/\sigma$\\ \hline
`quarks' & 224 & {\bf 262.9} 259.1 (261.2) & {\bf {}.038} {}.054 (.044) & {\bf {2.1}} 2.0 (2.0)\\
`LFU'  & 23 & {\bf 17.1} 39.4 (39.4) & {\bf {}.80} {}.018 ({}.018) & {\bf 0.2} 2.4 (2.4) \\ \hline
combined & 247 & {\bf 280.0} 298.5 (300.7) & {\bf {}.073} {}.014 (.011) & {\bf   1.8} 2.5 (2.5)\\
    \hline
  \end{tabular}
  \caption{\label{tab:SMpvals} SM quality-of-fit as calculated by \smelli{}
    with an updated $\overline{BR}(B_s\rightarrow \mu^+\mu^-)$ constraint (see
    \S\ref{sec:bsmm}). Only the values in bold include the 2022 LHCb 
    measurements of $R_{K^{(\ast)}}$~\cite{LHCb:2022qnv} and values in 
    parentheses are obtained \emph{without} updating the
    $\overline{BR}(B_s\rightarrow \mu^+\mu^-)$ constraint.
    $n_{obs}$ shows the number of observables in each category. $\chi^2_{SM}$ denotes 
    the $\chi^2$ statistic within each category, $p$ is the $p-$value of the
    category, and $s$ is the equivalent two-sided `number of $\sigma$' away
    from the     central value of a univariate normal distribution.
    The category  `LFU' contains lepton flavour universality violating
    flavour changing observables such as $R_K^{(\ast)}$, where theoretical
    uncertainties are relatively small. 
    The `quarks' category contains other flavour-changing $b$ observables, some of which
    have large theoretical uncertainties (which are nevertheless taken
    into account in the 
    calculation of $\chi^2$). 
  }
 \end{center}
  \end{table}
We display the current state-of-play as regards the tensions in
Table~\ref{tab:SMpvals}, as well as the change from including the LHCb
  measurements of $R_{K^{(\ast)}}$; the upshot is that the SM currently possesses a
1.8$\sigma$ tension with 247 combined measurements of $b\rightarrow s$
flavour-changing observables.
The December 2022 LHCb measurements have decreased this from 2.5$\sigma$.
We note here that the preselected
247 measurements
include several observables which do not involve the $(\bar b s) (\bar \mu
\mu)$ 
effective vertex and which also do not display large tensions, diluting
(i.e.\ raising) the
$p-$value and therefore decreasing $s/\sigma$.
In an analysis which concentrates more specifically on the fewer
\bsll\ anomaly observables on data in 2021, an 
estimate of $s=4.3\sigma$ was made~\cite{Isidori:2021vtc}.
We parenthetically observe from the table that in the context of
our SM global fit to many observables,
the update to the $\overline{BR}(B_s \rightarrow
\mu^+ \mu^-)$ constraint has only a small effect on the quality-of-fit. 

Prior to December 2022, it was well known in the
literature that two-parameter new physics models
can decrease $\chi^2$ by some
30 or so units~\cite{Alguero:2021anc,Altmannshofer:2021qrr,Ciuchini:2020gvn,Hurth:2021nsi} resulting in a significantly better fit than the SM\@. 
One can take the two parameters to be 
weak effective theory (WET) Wilson coefficients (WCs), defined via the WET
Hamiltonian density
\begin{equation} \label{eq:WET}
{\mathcal H}_{\mathrm{WET}} = - \frac{4 G_F}{\sqrt{2}} \sum_i (C_i^{\text{SM}} + C_i) \mathcal{O}_i + \text{H.c.},
  \end{equation}
where $C_i$ here denotes the \emph{beyond} the SM contribution to the WC,
$G_F$ is the Fermi decay constant and
$C_i^{\text{SM}}$ denotes the SM contribution to the WC\@.
The two dimension-6 operators $\mathcal{O}_i$
that can best ameliorate the \bsll\ fits are~\cite{Alguero:2021anc,Altmannshofer:2021qrr,Ciuchini:2020gvn,Hurth:2021nsi}
\begin{align}
  \mathcal{O}_9 &= \frac{e^2}{16 \pi^2} \left(\bar s \gamma_\mu P_L b\right)
  \left( \bar \mu \gamma^\mu \mu \right), 
\nonumber \\
  \mathcal{O}_{10} &= \frac{e^2}{16 \pi^2} \left(\bar s \gamma_\mu P_L b\right)
  \left( \bar \mu \gamma^\mu \gamma_5 \mu \right), \label{o9and10}
\end{align}
where $P_L$ is a left-handed projection operator in spinor space and $e$ is
the electromagnetic gauge coupling. Here $s$, $b$ and $\mu$ are 4-component Dirac
spinor fields of the strange quark, the bottom quark and the muon,
respectively. In Ref.~\cite{Greljo:2022jac} (Fig.~1), it was shown that the LHCb 2022 reanalysis of $R_{K^{(\ast)}}$ introduces a mild $1-2\sigma$ tension between the `quarks' category and `LFU' category of observable, if they are interpreted only in terms of $C_9$ and $C_{10}$. Despite this, we find that models which predict non-zero $C_9$ and $C_{10}$ still provide an improved fit to the combination of current flavour data as compared to the SM.

There
are two categories of beyond the SM fields
that can explain the \bsll\
anomalies at tree level in quantum field theory: leptoquarks, which are colour triplet scalar or vector bosons (with various possible electroweak quantum numbers),
and $Z^\prime$s, which are SM-neutral vector bosons. In each category, the new state must have family non-universal
interactions, coupling to $\overline{b_L}$ and $s_L$.
The observables in which the measurements are in tension with SM
  predictions involve di-muon  pairs. The new physics state should therefore
  certainly
  couple to di-muon pairs, but in order to agree with the December 2022 LHCb
  measurements $R_K$ and $R_{K^\ast}$, one may also couple it to di-electron
  pairs with a similar strength, although then the scenario may become
  strongly constrained by LEP constraints~\cite{Greljo:2022jac}.
  For the coupling to di-muon pairs, fits to flavour data by different groups
  agree that the new physics state should couple to \emph{left}-handed
  di-muons\footnote{Our  
  convention here is that `left-handed di-muons' signifies both $\mu_L:= P_L
  \mu$ and  $\overline{\mu_L}$.}. 
There is a preference from the data for a coupling
to muons and
fits to flavour data by different groups agree
that the new physics state should couple to left-handed di-muons\footnote{Our
  convention here is that `left-handed di-muons' signifies both $\mu_L:= P_L \mu$ and
  $\overline{\mu_L}$.}.
(A purely left-handed coupling to muons realises the limit
$C_9=-C_{10}$, which will be a prediction of the benchmark
leptoquark model that we examine.)
However, depending upon the treatment of the predictions of 
observables in the `quarks' 
category, the different fits may or may not have a mild
preference~\cite{Egede:2022rxc}  
for the new physics field to couple
to right-handed di-muons in addition
with a similar strength to the coupling to left-handed
di-muons (e.g.\ realising the limit where $C_9\neq 0$ but $C_{10}=0$;
this will be the prediction of the benchmark $Z^\prime$ model which we shall
choose)~\cite{Allanach:2020kss}. 
\begin{figure}
  \begin{center}
    \includegraphics[width=210pt]{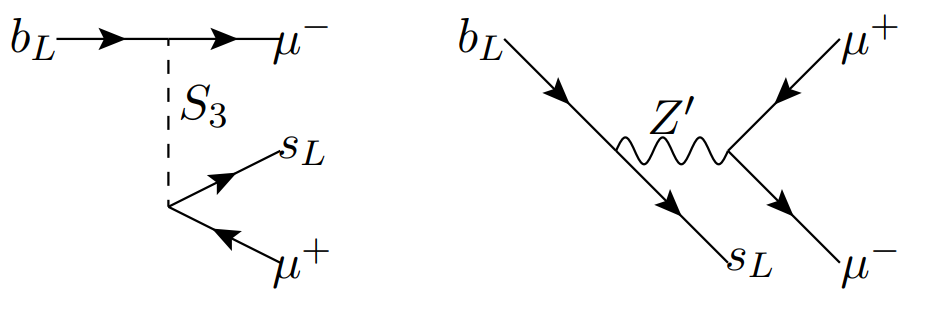}
\caption{\label{fig:NCBAs} Tree-level Feynman diagrams of the new physics contributions to the \bsll\ observables in a suitable
  scalar $S_3$ leptoquark model (left panel) 
  and in a suitable $Z^\prime$ model (right panel).}
\end{center}
\end{figure}
We display example 
Feynman diagrams contributing to $b\rightarrow s \mu^+\mu^-$ processes for a
scalar leptoquark and for a $Z^\prime$
in Fig.~\ref{fig:NCBAs}.

On the model building side,
leptoquark explanations for the \bsll\ anomalies have received a lot of
attention. For scalar leptoquarks, only the $S_3 \sim {\bf
  (\overline{3},3)}_{1/3}$ leptoquark can fit the \bsll\ anomaly
data~\cite{Hiller:2017bzc,Angelescu:2018tyl,Angelescu:2021lln}. Theoretical
challenges include explaining why the scalar leptoquark should be as light as
the TeV scale, which can be addressed in the partial compositeness
framework~\cite{Gripaios:2009dq,Gripaios:2014tna}, and explaining why the
leptoquark does not have baryon-number violating or lepton flavour violating
(LFV) couplings. The latter can be achieved by
gauging an additional symmetry~\cite{Davighi:2020qqa,Greljo:2021xmg,Greljo:2021npi,Davighi:2022qgb}, extending previous work using (global) flavour symmetries~\cite{deMedeirosVarzielas:2015yxm,Heeck:2022znj}. 
For vector leptoquarks, a $U_1 \sim {\bf (3,1)}_{2/3}$ or a $U_3 \sim {\bf
  (3,3)}_{2/3}$ leptoquark can significantly ameliorate the quality-of-fit of
the SM~\cite{Angelescu:2018tyl,Angelescu:2021lln}, but a simplified model is
non-renormalisable and so a ultra-violet (UV) complete model is needed. In the case of the $U_1$ leptoquark, a suitable class of extensions of the SM gauge group based on the `4321' gauge group $SU(4)\times SU(3)\times SU(2)\times U(1)$ was invented~\cite{DiLuzio:2017vat,DiLuzio:2018zxy,Greljo:2018tuh}, which has inspired much further work. A family-triplet leptoquark model can successfully fit all data including the December 2022 LHCb $R_K$ measurements~\cite{Greljo:2022jac}.

Similarly, many
$Z^\prime$ models have been constructed which also significantly improve
the SM's poor quality-of-fit. Many are based on spontaneously broken anomaly-free $U(1)_X$
gauge extensions of the SM that yield such a $Z^\prime$, for various choices of $X$. Examples include gauged muon minus tau
lepton
number $X=L_\mu-L_\tau$~\cite{Altmannshofer:2014cfa,Crivellin:2015mga,Crivellin:2015lwa,Crivellin:2015era,Altmannshofer:2015mqa}, for which the required $\ZP$ coupling to $\bar{s}_L b_L$ is mediated via mixing with heavy vector-like quarks, 
third family baryon number minus muon lepton
number $X=B_3-L_2$~\cite{Alonso:2017uky,Bonilla:2017lsq,Allanach:2020kss},
third family
hypercharge $X=Y_3$~\cite{Allanach:2018lvl,Davighi:2019jwf,Allanach:2019iiy,Allanach:2021kzj}, and other
assignments~\cite{Sierra:2015fma,Celis:2015ara,Greljo:2015mma,Falkowski:2015zwa,Chiang:2016qov,Boucenna:2016wpr,Boucenna:2016qad,Ko:2017lzd,Alonso:2017bff,Tang:2017gkz,Bhatia:2017tgo,Fuyuto:2017sys,Bian:2017xzg,King:2018fcg,Duan:2018akc,Kang:2019vng,Calibbi:2019lvs,Altmannshofer:2019xda,Capdevila:2020rrl,Davighi:2021oel,Allanach:2022bik}. An attempt to fit the post-December 2022 data using a lepton-flavour-universal $3B_3-L$ data was made in Ref.~\cite{Greljo:2022jac}, but the model was found to be in significant tension with LEP Drell-Yan constraints coming from $e^+ e^- \to \ell^+ \ell^-$.

\begin{figure}
\begin{center}
\includegraphics[width=240pt]{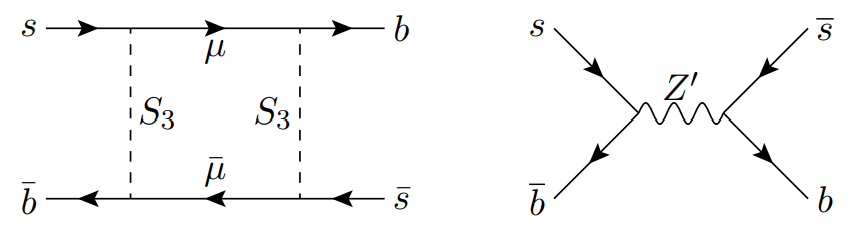}
  \caption{\label{fig:BsBsbar} Example leading order Feynman diagrams of beyond
  the SM 
  contribution to $B_s - \overline{B_s}$ mixing via the $S_3$
  leptoquark (left panel) or the \BthreeLtwo{} (right panel).}
\end{center}
\end{figure}
  The primary aim
    of this paper
  is to compare global fits to current data of a leptoquark model with
  those of a $Z^\prime$ model.
  We pick two
    simple and comparable benchmark models that have been studied in previous literature. 
They shall be defined in
    the next Section.
While current tensions in semi-leptonic \bsll\ transitions 
between measurements and SM predictions can be
ameliorated by either a leptoquark or a $\ZP$,
it is anticipated
that the $\Delta M_s$ observable~\cite{King:2019lal}, which parameterises $B_s - \overline {B_s}$ mixing 
and does not exhibit a significant tension with the SM prediction, 
has the power to discriminate between leptoquark and $Z^\prime$ explanations. 
Because the $\ZP$ model predicts \emph{tree-level}
contributions to $\Delta M_s$, while
the dominant contributions from the leptoquark
are at one-loop order (see 
Fig.~\ref{fig:BsBsbar}) and therefore suppressed, one generically expects the constraints from $\Delta M_s$ to be stronger for
$Z^\prime$ models. However, the extent of the impact of this
observable 
on the fits should be evaluated quantitatively
in each model for comparison.
A secondary goal of the present paper is to examine the impact of the
December 2022 LHCb measurements\footnote{We shall use these
measurements in our default fits, but in key cases, display the change that
they induce.} of $R_{K^{(\ast)}}$.

For both the leptoquark and the $\ZP$, we choose to focus on TeV-scale masses
so that the models can avoid the bounds originating from direct searches at the LHC\@. 
For TeV-scale masses, the new physics state's contribution to the anomalous magnetic
moment of the muon $(g-2)_\mu$ is negligible without augmenting the model
  with additional fields~\cite{Allanach:2015gkd}
and so we exclude this observable
from our fits. 

In each model, we do expect there to be
additional fields that we (usually) shall assume have a negligible
effect upon the
\bsll\ observables either because they
are 
too weakly coupled or because they are too heavy. For example, additional
heavy fermionic fields are expected in order to generate SM-fermion mixing. If additional fermionic fields are added in
vector-like representations of the gauge group\footnote{We shall refer to
    such fermions, following common practice, as `vector-like fermions'.}, then the model remains anomaly
free and large masses for said fermions are allowed by the gauge symmetry.
Additional fields could be added in order to explain the current tension in
$(g-2)_\mu$, for example additional vector-like 
leptons~\cite{Allanach:2015gkd}, where a one-loop diagram with a $Z^\prime$
along with additional leptons in the loop can resolve it. Other
heavy fields associated with dark matter could also be added.
Here, our effective quantum field theory is supposed to be valid at and below
the scale of the mass of the $Z^\prime$ or leptoquark state, i.e.\ the TeV
scale, and we  
assume that the effect of other heavy fields on the pertinent
flavour phenomenology is negligible. 

The parameter space of the \BthreeLtwo\ model was adapted
  to \bsll\ data and various phenomenological constraints were applied
  in Refs.~\cite{Alonso:2017uky,Bonilla:2017lsq,Allanach:2020kss,Allanach:2021gmj}. 
  Some more recent flavour data and LHC $\ZP$ search limits were employed
  to constrain the model in Ref.~\cite{Azatov:2022itm}.
  In that reference, similar constraints were also
  applied to the $S_3$ leptoquark model (estimated before with earlier
  data by Refs.~\cite{Allanach:2017bta,Allanach:2019zfr}), before calculating future
  collider sensitivities of the two models. In all of these prior works, the parameters of each model
  were fit to central values for $C_9$ and $C_{10}$ obtained from fits to $b\to s \ell^+ \ell^-$
  data, assuming that all other SM effective field theory (SMEFT) operators were zero\footnote{In
    Ref.~\cite{Gherardi:2020qhc}, a global fit of the $S_3$ model to
    \bsll\ \emph{and} $b 
    \rightarrow c \tau \nu$ anomalies was performed. Here, we neglect the
    $b      \rightarrow c \tau \nu$
    anomalies because the evidence for new physics effects in them is not
    as strong 
    as it is in the \bsll\ anomalies.}. The additional constraints, 
  including those from meson mixing and LHC searches,
  were individually applied to obtain 95$\%$ CL bounds.
  The bottom line of Ref.~\cite{Azatov:2022itm} relevant for the present
    paper 
  is that, in each model, there
  is a sizeable parameter space that evades current search limits
  and satisfies other phenomenological constraints, and that simultaneously
  explains the \bsll\ anomalies. 
  In the present paper we go beyond the prior studies by performing a
  side-by-side global flavour fit of each model to identical empirical datasets. All dimension-6 SMEFT 
  operators are included -- not just those that generate $C_9$ and $C_{10}$
  -- including 4-quark and 4-lepton operators induced at tree-level in the $\ZP$ model, and 
  those operator contributions generated by one-loop renormalisation effects.
  We then perform a goodness-of-fit test on each
  model, comparing the best-fit points to quantify any statistical preference
  of the data for either model. 

The paper proceeds as follows: \S\ref{sec:models}, we
introduce the $S_3$ leptoquark model and the \BthreeLtwo. In
order to make further progress, some assumptions about fermion mixing must be
made and we lay these out first; they are identical for each of the two
models. In 
\S\ref{sec:SMEFT} we present 
the matching to the WCs of dimension-6 operators in the 
SMEFT resulting from integrating out the new physics state.
The fit to each model will be described by 2 effective parameters extra to the
SM\@. For given values of these parameters, the dimension-6 SMEFT operator WCs are then given and can be passed as input to \smelli{}, which calculates the
observables and quality-of-fit. Fit results are presented in
\S\ref{sec:fits}, where we shall observe that 
the leptoquark model has a similar quality-of-fit as the $Z^\prime$ model,
indicating that the $\Delta 
M_s$ observable was less discriminating than one might have expected. We
summarise and conclude in \S\ref{sec:conc}.

\section{Models\label{sec:models}}
It is our purpose here to add a TeV-scale new physics field to the SM with
couplings that facilitate an explanation of the \bsll\
anomalies. We will not develop the model-building fully into the UV,
being content to 
describe the important phenomenology such that experimental assessments
of the \bsll\ anomalies
can be made. 
We begin by writing down our conventions as regards fermion mixing, since
they are
common to both the $S_3$ model and the \BthreeLtwo.

\subsection{Fermion mixing conventions}
In the fermionic fields' gauge eigenbasis, which we indicate via `primed' symbols, we write 
\begin{eqnarray}
{\bf u_L'}&=&\left( \begin{array}{c} u_L' \\ c_L' \\ t_L' \\ \end{array}
\right), \qquad
{\bf d_L'}=\left( \begin{array}{c} d_L' \\ s_L' \\ b_L' \\ \end{array}
\right), \qquad
{\bf e_L'}=\left( \begin{array}{c} e_L' \\ \mu_L' \\ \tau_L' \\ \end{array}
\right), \qquad
{\bm \nu_L'}=\left( \begin{array}{c} {\nu_e'}_L \\ {\nu_\mu'}_L
  \\ {\nu_\tau'}_L \\ \end{array} \right),
 \nonumber \\ 
{\bf u_R'}&=&\left( \begin{array}{c} u_R' \\ c_R' \\ t_R' \\ \end{array}
\right), \qquad
{\bf d_R'}=\left( \begin{array}{c} d_R' \\ s_R' \\ b_R' \\ \end{array}
\right),\qquad
{\bf e_R'}=\left( \begin{array}{c} e_R' \\ \mu_R' \\ \tau_R' \\ \end{array}
\right), \qquad
{\bm \nu_R'}=\left( \begin{array}{c} {\nu_e'}_R \\ {\nu_\mu'}_R
  \\ {\nu_\tau'}_R \\ \end{array} \right),
\end{eqnarray}
along with the SM fermionic electroweak doublets
\begin{equation}
{\bf Q'}_i=\left( \begin{array}{c} {\bf u_L'}_i \\ {\bf d_L'}_i \end{array}
\right),\qquad
{\bf L'}_i=\left( \begin{array}{c} {\bm \nu_L'}_i \\ {\bf e_L'}_i \end{array}
\right).  
\end{equation}
The SM fermions
acquire masses after the SM Brout-Englert-Higgs mechanism through 
\begin{eqnarray}
-\mathcal{L}_{Y}&=&\overline{\bf Q'}  Y_u \tilde H {\bf u'_R} +
\overline{\bf Q'}  Y_d H  {\bf d'_R} +
\overline{\bf L'}  Y_e H  {\bf e'_R} + 
\overline{\bf L} Y_\nu \tilde H {\bm \nu'_R} + 
\frac{1}{2}{\overline{{\bm \nu_R'}^c}} M 
  {\bm \nu_R'} + H.c. , 
 \label{yuk}
\end{eqnarray}
where $Y_u$, $Y_d$ and $Y_e$ are dimensionless complex coupling constants,
each written as a 3 by 3 matrix in family space. 
For both models that we
will consider, some  of 
these Yukawa couplings should be interpreted as effective
  dimension-4 operators that arise in the SMEFT limit, once heavier degrees of
  freedom are integrated out; see \S\ref{sec:mixing} for details.
Gauge indices have been
omitted in (\ref{yuk}).
The matrix $M$ is a 3 by 3 complex symmetric matrix of mass dimension
1, ${\Phi}^c$ denotes the
charge conjugate of a field $\Phi$ and $\tilde H := ({H^0}^\ast, -H^-)^T$. 

We may write
$H=(0,\ (v + h)/\sqrt{2})^T$ after electroweak symmetry breaking,
where 
$h$ is the physical Higgs
boson field
and (\ref{yuk}) includes the fermion mass terms 
\begin{eqnarray}
-\mathcal{L}_{Y}&=&\overline{\bf u'_L} V_{u_L} V_{u_L}^\dagger m_u V_{u_R}
V_{u_R}^\dagger {\bf u'_R} + 
\overline{\bf d'_L} V_{d_L} V_{d_L}^\dagger m_d  V_{d_R} 
V_{d_R}^\dagger {\bf d'_R} + 
\overline{\bf e'_L} V_{e_L} V_{e_L}^\dagger m_e  V_{e_R} 
V_{e_R}^\dagger {\bf e'_R} + \nonumber \\ &&
\frac{1}{2} ( {\overline{\bm \nu_L'}}\ \overline{{\bm \nu_R'}^c}) M_\nu
\left( \begin{array}{c} {\bm {\nu_L'}}^c \\ {\bm \nu_R'} \\
\end{array}
  \right)
  +H.c. + \ldots, \label{diracMass}
\end{eqnarray}
where
\begin{equation}
M_\nu = \left( \begin{array}{cc} 0 & m_{\nu_D} \\
  m_{\nu_D}^T & M \\ \end{array} \right),
\end{equation}
$V_{I_L}$ and $V_{I_R}$ are 3 by 3 unitary mixing matrices for each
field species $I$, 
$m_u:=v Y_u/\sqrt{2}$, $m_d:=v
Y_d/\sqrt{2}$, $m_e:=v Y_e/\sqrt{2}$ and $m_{\nu_D}:=v Y_\nu/\sqrt{2}$.
The final explicit term in (\ref{diracMass})
incorporates the see-saw
mechanism via a 6 by 6 complex
symmetric mass matrix. Since the elements in $m_{\nu_D}$ are much smaller than
those in $M$, we perform a rotation to obtain a 3 by 3 complex symmetric
mass matrix for the three light neutrinos. These approximately
coincide with the left-handed weak eigenstates ${\bm \nu'_L}$, whereas
three heavy neutrinos approximately correspond to the right-handed weak
eigenstates ${\bm \nu'_R}$. The neutrino mass term of (\ref{diracMass}) becomes, to a good
approximation, 
\begin{equation}
-  {\mathcal L}_{\nu} =
  \frac{1}{2} {\overline {\bm \nu_L'^c}} m_\nu {\bm \nu_L'} +
\frac{1}{2} {\overline {\bm \nu_R'^c}} M {\bm \nu_R'} + H.c., 
  \end{equation}
where $m_\nu:= m_{\nu_D}^T M^{-1} m_{\nu_D}$ is a complex symmetric 3 by 3
matrix. 

Choosing
$V_{I_L}^\dagger m_I  V_{I_R}$ to be diagonal, real and positive for $I
\in \{ u,d,e\}$, and
$V_{{\nu}_L}^T m_\nu  V_{{\nu}_L}$ to be diagonal,
real and positive 
(all in ascending order of mass
from the top left toward the bottom right of the matrix), we can identify the 
{\em non}-primed {\em mass}\/ eigenstates\footnote{${\bf P}$ and ${\bf P}'$ are column vectors.}
\begin{equation}
  {{\bf P}} = V_P^\dag {\bf P'} \ \text{where}\ {P} \in \{{u_R},\ {
    d_L},\ { u_L},\ { e_R},\ { u_R},\ { d_R},\ {\nu}_{L},\ {
    e_L}\}. \label{fermion_rotations} 
  \end{equation}
We may then find the CKM matrix $V$ and the
Pontecorvo-Maki-Nakagawa-Sakata (PMNS) matrix $U$ in terms of the fermionic
mixing matrices:
\begin{equation}
V=V_{u_L}^\dagger V_{d_L}, \qquad U = V_{\nu_L}^\dagger V_{e_L}. \label{mix}
\end{equation}

\subsection{Fermion mixing ansatz}

To make phenomenological progress with our models, we shall need to fix the
$V_{P}$. We make some fairly strong assumptions about these 3 by 3
unitary matrices, picking a simple ansatz which is not immediately ruled out
by 
strong flavour changing neutral current constraints on charged lepton flavour
violation or neutral current flavour violation in the first two families of
quark. Thus, we pick $V_{e_R}=V_{d_R}=V_{u_R}=V_{e_L}=I$, the 3 by 3 identity
matrix.
A non-zero $(V_{d_L})_{23}$ matrix element is required for both the $S_3$
model and the \BthreeLtwo\ to mediate new physics contributions to
$b\rightarrow s \mu^+\mu^-$ transitions.
We capture the important quark mixing (i.e.\ that
between $s_L$ and $b_L$) in $V_{d_L}$ as
\begin{equation}
  V_{d_L} = \left(\begin{array}{ccc} 1 & 0 & 0 \\
  0 & \cos \theta_{23} & \sin \theta_{23} \\
  0 & -\sin \theta_{23} & \cos \theta_{23} \\ \end{array}
  \right). \label{ansatz}
\end{equation}
$V_{\nu_L}$ and $V_{u_L}$ are fixed
by (\ref{mix}), where we use the experimentally determined values for
the entries of $V$ and $U$
via the central values in
the standard parameterisation from
Ref.~\cite{ParticleDataGroup:2020ssz}.
Having fixed all of the fermionic mixing matrices, we have provided an
ansatz that could be perturbed
around for a more 
complete characterisation of the models. We leave such perturbations aside for
the present paper.

Having set the conventions and an ansatz for the fermionic mixing matrices,
we now introduce
the $S_3$ leptoquark model and the \BthreeLtwo\ in turn.

\subsection{Benchmark $S_3$ leptoquark model}

For a bottom-up leptoquark model to explain the \bsll\ anomalies we consider a
scalar leptoquark rather than a vector leptoquark, since inclusion of the
latter would require an extension of the SM gauge symmetry (which necessarily
requires more fields beyond the leptoquark). 
To fit the \bsll\ anomalies alone, the
$S_3 \sim (\bar{\bf 3}, {\bf 
  3})_{1/3}$ leptoquark is a good candidate because it couples  to
left-handed quarks (and left-handed muons), which can broadly agree with
\bsll\ data as described in \S\ref{sec:intro}.

The Lagrangian density includes the following extra terms due to the $S_3$
leptoquark: 
\begin{align}
{\mathcal L}_{S_3} &= |D_\mu S_3|^2 - M_{S_3}^2 |S_3|^2 - \lambda_{H3}|H|^2|S_3|^2 - \lambda_{\epsilon H3} i\epsilon^{abd} (H^\dagger \sigma^a H) S_3^{b\dagger} S_3^d - V(S_3) \nonumber \\
&+ \left[\lambda_{xy} \overline{{{Q_x}^\prime}^c}
  (i\sigma^2)
  \sigma^a S_3^a L_y^\prime +
  \kappa_{xy} \overline{{Q^\prime_x}^c}
   (i\sigma^2)
  (\sigma^a S_3^a)^{\dagger}
Q_y^\prime + \text{H.c.} \right], \label{eq:S3lag}
\end{align}
where $x,y \in \{1,2,3\}$ are family indices, $a,b,d \in \{1,2,3\}$ are
$SU(2)_L$ adjoint indices (other gauge and spinor indices are all suppressed) and $\sigma^a$
are Pauli matrices for $SU(2)_L$.
Here, $\lambda_{\epsilon H3}, \lambda_{H3}
\in \mathbb{R}$. 
By charging the leptoquark under a lepton-flavoured $U(1)_X$ gauge symmetry~\cite{Davighi:2020qqa,Greljo:2021xmg,Greljo:2021npi,Davighi:2022qgb} one can ban the proton-destabilising di-quark operators
($\kappa_{xy} \to 0$), and furthermore pick out only couplings of the
leptoquark that
are muonic: 
\begin{equation}
  \lambda_{xy} = \lambda \delta_{y 2} \gamma_x \, , \qquad \gamma \in \C^3, \,\ \gamma_x \gamma_x^\ast = 1,\,\ \lambda \in \mathbb{R}\, .
  \label{limits}
\end{equation}
We shall assume (\ref{limits}).
The leptoquark's couplings to fermions are now specified by a unit-normalised complex 3-vector
$\gamma$ in quark-doublet family space. The factor of $\lambda$ fixes the overall
strength of the 
interaction of the leptoquark with the SM fermions. 

We shall moreover assume that $\gamma=(0,0,1)$,
which could 
be enforced with further family symmetries, for example by an approximate
$U(2)_Q$ global symmetry, or by a particular choice of anomaly-free gauged
$U(1)_X$ that is quark non-universal. The space of anomaly-free  solutions
  $U(1)_X$ symmetries with these properties was explored systematically in
  Ref.~\cite{Greljo:2021npi} (see Sec 2.3), building on the results of
  Ref.~\cite{Allanach:2018vjg}.
For concreteness, we here choose to gauge
\begin{equation} \label{eq:muoquark_X}
X = B_3 + L_2 - 2L_3,
\end{equation}
with an $S_3$ charge of $X_{S_3}=-2$. 
This $U(1)_X$ symmetry allows the coupling to
$\overline{{Q_3^\prime}^c} L_2^\prime$ but no other quark-lepton pairs, and moreover
bans the coupling to di-quark operators.
We shall discuss the Yukawa sector in \S\ref{sec:mixing}.

We now perform a global rotation in family space on $Q_i^\prime$ such that
the $d_{L_i}$ fields are in their mass basis: $Q_i=(V_{d_L}^\dag
Q')_i=((V^\dag u_L)_i,\ d_{L_i})$, whereas in a similar fashion, $L_i$ 
already have the $e_{L_i}$ 
fields in their mass basis: $L_i=((U^\dag \nu_L)_i,\ e_{L_i})$. The $S_3$
leptoquark's couplings to fermionic fields become
\begin{align}
  {\mathcal L}_{S_3} =&\ldots+
  \lambda \overline{{\bf Q}^c}_x (V_{d_L}^T)_{x3} (i\sigma^2) \sigma^a S_3^a L_2    
  + H.c.,
\end{align}
and since $(V_{d_L})_{23} \neq 0$ and $(V_{d_L})_{33}\neq 0$ for
$\theta_{23}\neq 0$ in (\ref{ansatz}), the model possesses the correct
couplings to mediate \bsll\ transitions as depicted in the left-hand panel of
Fig.~\ref{fig:NCBAs}.

\subsection{Benchmark \BthreeLtwo} \label{sec:ZPmodel}

We extend the SM by a $U(1)_X$ gauge group and a SM-singlet complex scalar $\theta$, with field charges as in
Table~\ref{tab:fields}.
\begin{table}
\begin{equation*}
\begin{array}{|c|ccc|ccc|ccc|ccc|ccc|ccc|c|c|}
\hline
& Q_1^\prime & Q_2^\prime & Q_3^\prime & u_1^\prime& u_2^\prime & u_3^\prime & d_1^\prime & d_2^\prime & d_3^\prime & L_1^\prime & L_2^\prime &
L_3^\prime &e_1^\prime & e_2^\prime &e_3^\prime & \nu_1^\prime &\nu_2^\prime &\nu_3^\prime & H & \theta \\ \hline
 SU(3) && {\bf 3} &  &  & {\bf 3} &  & & {\bf 3} &  & & {\bf 1} &  & & {\bf 1}
 & &  & {\bf 1} & & {\bf 1} & {\bf 1}\\
 SU(2) & & {\bf 2} & &  & {\bf 1} &  & & {\bf 1} & & & {\bf 2} & & & {\bf 1} &
 & & {\bf 1}  & & {\bf 2} & {\bf 1} \\
 U(1)_Y & & 1 &  &  & 4 &  & & -2 & & & -3 & & & -6 & &  & 0 & & 3 & 0\\ 
\hline
 U(1)_{X} & 0 & 0 & 1 & 0 & 0 & 1 & 0 & 0 & 1 & 0 & -3 & 0 & 0 & -3 & 0
 & 0 & -3 & 0 & 0 & q_\theta\\
\hline
\end{array}
\end{equation*}
\caption{Representations of fields under the SM gauge factors, which
  are family universal, together with their representations under the
  family non-universal gauged 
  $U(1)_X$ symmetry on which our $\ZP$ model is based. We use the minimal
  integer normalisation for the charges under each $U(1)$ factor and
  we shall specify $q_\theta \in \{-1,\ 1\}$. 
  All fields are Weyl fermions except for the complex scalar Higgs doublet $H$
  and the complex scalar flavon $\theta$.
 \label{tab:fields}
}
\end{table}
This is the ${B_3-L_2}$ $Z^\prime$ model of Ref.~\cite{Allanach:2020kss} that
provided a simple bottom-up description of the models in
Refs.~\cite{Alonso:2017bff,Bonilla:2017lsq} (these possess additional fields),
which was shown to explain the \bsll\ anomalies.
Note that we neglect any effects coming from
  $U(1)_X-U(1)_Y$  
mixing. This approximation may be motivated (at tree-level) by further model
building, for example by embedding each $U(1)$ Lie algebra generator in  the
Cartan subalgebra of some 
semi-simple gauge group which subsequently breaks to $SU(3)\times SU(2)_L
\times U(1)_Y \times U(1)_X$.\footnote{Semi-simple gauge extensions in which the $B_3-L_2$ model embeds can 
be found using Ref.~\cite{Davighi:2022dyq}. Such extensions can evade
  bounds coming from requiring a lack of Landau poles~\cite{Bause:2021prv}.}
The mixing would be set to zero at the scale of
this breaking and then generated at one-loop order by running down to the mass scale
of the $\ZP$. The resulting mixing is small unless the two scales are
separated by a large hierarchy.

The $U(1)_{X}$ symmetry is broken by $\langle \theta \rangle := v_X /
\sqrt{2}\sim {\mathcal O}(\text{TeV})$ and so the
$Z^\prime$ acquires a tree-level mass 
\begin{equation}
  M_{Z^\prime} = g_{Z^\prime} v_X, \label{eq:MZp}
\end{equation}
where $g_{Z^\prime}$ is the
$U(1)_X$ gauge coupling. 
The couplings of the $Z^\prime$ boson are then 
\begin{equation}
{\mathcal L}_{Z^\prime \psi} =
-g_{Z^\prime} \left( 
{\overline{{Q_3'}}} \slashed{Z}^\prime {Q_3'} +
{\overline{{u_3'}_R}} \slashed{Z}^\prime {u_3'}_R +
{\overline{{d_3'}_R}} \slashed{Z}^\prime {d_3'}_R -3
{\overline{{L_2'}}} \slashed{Z}^\prime {L_2'}-3
{\overline{{e_2'}_R}} \slashed{Z}^\prime {e_2'}_R-3
{\overline{{\nu_2'}_R}} \slashed{Z}^\prime {\nu_2'}_R
\right), \label{Zpcouplings}
\end{equation}
in the primed gauge eigenbasis.


Re-writing (\ref{Zpcouplings}) in the mass basis, we obtain the $\ZP$
couplings to the fermionic fields
\begin{eqnarray}
{\mathcal L}_{Z^\prime \psi} &=&-
g_{Z^\prime}  \left(
{\overline{{\bf Q}}} \Lambda^{(d_L)}_\Xi \slashed{Z}^\prime {\bf Q} 
+
{\overline{{\bf u_R}}} \Lambda^{(u_R)}_\Xi \slashed{Z}^\prime {\bf u_R} +
{\overline{{\bf d_R}}} \Lambda^{(d_R)}_\Xi \slashed{Z}^\prime {\bf d_R}
\right. \nonumber \\ &&
\left.-3 {\overline{{\bf L}}} \Lambda^{(e_L)}_\Omega \slashed{Z}^\prime {\bf L} 
- 3 {\overline{{\bf e_R}}} \Lambda^{(e_R)}_\Omega \slashed{Z}^\prime {\bf e_R}
-3 {\overline{{\bm \nu_R}}} \Lambda^{(\nu_R)}_\Omega \slashed{Z}^\prime {\bm \nu_R}
\right), \label{Zpcoupmass}
\end{eqnarray}
where we have 
defined 
\begin{equation}
\Lambda^{(P)}_\beta := V_{P}^\dagger \beta V_{P} , \qquad \beta \in \{ \Xi, \Omega \}
\label{lambdas}
\end{equation}
and
\begin{equation}
\Xi := \left(\begin{array}{ccc}
0 & 0 & 0 \\ 0 & 0 & 0 \\ 0 & 0 & 1 \\
\end{array}\right), \qquad
\Omega := \left(\begin{array}{ccc}
0 & 0 & 0 \\ 0 & 1 & 0 \\ 0 & 0 & 0 \\
\end{array}\right). \qquad
\end{equation}
For $\theta_{23}\neq 0$, (\ref{lambdas}) implies that the diagram in the
right-hand panel of Fig.~\ref{fig:NCBAs} contributes to the
\bsll\ observables. We note here that the $\overline{s_L} \gamma_\mu Z^\mu b_L$
coupling 
is proportional to the gauge coupling multiplied by
\begin{equation}
  {\Lambda_\Xi^{(d_L)}}_{23} = - \frac{1}{2} \sin 2\theta_{23} \label{angle}
\end{equation}
from (\ref{ansatz})
  and (\ref{lambdas}).

\subsection{The origin of quark mixing} \label{sec:mixing}

The flavoured $U(1)_X$ symmetries that we gauge do not allow the complete set of Yukawa couplings at
the renormalisable level, for either benchmark model.
For both models,  the following quark Yukawa textures are populated at
dimension-4: 
\begin{equation} \label{eq:ren-quark-yuk}
Y_u\sim \begin{pmatrix}
\times&\times&0\\
\times&\times&0\\
0&0&\times\\
\end{pmatrix}, \qquad 
Y_d \sim \begin{pmatrix}
\times&\times&0\\
\times&\times&0\\
0&0&\times\\
\end{pmatrix}, 
\end{equation}
consistent with the $SU(2)_q \times SU(2)_u \times SU(2)_d$ accidental flavour
symmetry of the gauge sector (under which the light quarks transform as
doublets while the third family transform as
singlets~\cite{Kagan:2009bn,Barbieri:2011ci}). 
For the gauge symmetry (\ref{eq:muoquark_X}) in our leptoquark model, the charged lepton Yukawa matrix is strictly diagonal at dimension-4, 
while for the $B_3-L_2$ symmetry of \S\ref{sec:ZPmodel}, the dimension-4 charged lepton Yukawa matrix is:
\begin{equation}
Y_e \sim \begin{pmatrix}
\times&0&\times\\
0&\times&0\\
\times&0&\times\\
\end{pmatrix}\, .
\end{equation}
Either of these Yukawa structures for the charged leptons is sufficient to
  reproduce the observed charged lepton
  masses.

On the other hand,
in order to reproduce the non-zero CKM mixing angles between (left-handed) light quarks and third family quarks,
some of the zeroes in (\ref{eq:ren-quark-yuk}) must be populated by operators that are
  originally higher-dimension and that
  originate from integrating out more massive quantum fields.
These mixing angles are therefore expected to be small in this framework, in agreement with observations.

For either of our benchmark models, let us set the charge of the $U(1)_{X}$-breaking scalar field $\theta$ to be $q_\theta = 1$.
The remaining Yukawa couplings can then arise from dimension-5 operators, as
discussed explicitly in Ref.~\cite{Greljo:2021npi} (Section 2.3), 
\begin{equation} \label{eq:dim5yuk}
\mathcal{L} \supset 
\frac{C_U^i}{\Lambda_U} \overline{Q^\prime_i} \tilde H \theta^{\ast} u^\prime_{3 R} 
+ \frac{C_D^i}{\Lambda_D} \overline{Q^\prime_i} H \theta^{\ast} d^\prime_{3 R} 
+ \frac{\tilde{C}_U^i}{\Lambda_Q} \overline{Q^\prime_3} \tilde{H} \theta u^\prime_{i R} 
+ \frac{\tilde{C}_D^i}{\Lambda_Q} \overline{Q^\prime_3} H \theta
d^\prime_{i R} \, , 
\end{equation}
where in each term the index $i\in \{1,2\}$ runs over the light families, and
where $\Lambda_{U,D,Q}$ are effective scales associated with the
renormalisable new physics
responsible for these effective operators.
The first two operators on the
right-hand-side of (\ref{eq:dim5yuk}) set the 1-3 and 2-3 rotation angles for
{\em left}-handed quarks, which must be non-zero to account for the full CKM
matrix, while the second two operators would give rotation angles for {\em
  right}-handed fields, which are not required by data to be non-zero. 
If these right-handed mixing angles are approximately zero, then the
resulting $\ZP$ will couple only to left-handed not right-handed $\bar{s}b$
currents. (Sizeable $\ZP$ couplings to right-handed
$\bar{s}b$ currents are 
less favoured by the \bsll data~\cite{Aebischer:2019mlg}).

A simple UV origin for such operators, which can moreover predict the
  desirable relation $\tilde C_i^u \approx \tilde C_i^d \approx 0$, is to
  integrate out heavy vector-like quarks (VLQs). Since these
  VLQs also give other contributions to low-energy observables, notably to
  $B_s$ meson mixing which is a particular focus of this work, we expand upon
  this aspect of the UV set-up in a little detail.\footnote{We do not consider this discussion to be definitive of our model --- there
    could be other ways to provide the desired patterns --- but rather as an
    existence proof.}
  Consider adding a pair
  of VLQs in the representations 
\begin{equation} \label{eq:VLQreps}
\Psi_{L,R}^D \sim {\bf (3,1,} -2,0), \qquad \Psi_{L,R}^U \sim {\bf (3,1,} +4,0) 
\end{equation}
of the gauge group $SU(3)\times SU(2)_L \times U(1)_Y \times U(1)_X$. This
permits the following terms in the renormalisable Lagrangian, 
\begin{align} \
\mathcal{L} \supset 
&M_D \overline{\Psi_L^D} \Psi_R^D  
+ \lambda_D^i \overline{Q^\prime_i} H \Psi_R^D 
+ \kappa_D \overline{\Psi^D_L} \theta^\ast d^\prime_{3 R} \nonumber \\
+ &M_U \overline{\Psi_L^U} \Psi_R^U 
+ \lambda_U^i \overline{Q^\prime_i} \tilde{H} \Psi_R^U 
+ \kappa_U \overline{\Psi^U_L} \theta^\ast u^\prime_{3 R} + \text{~H.c.}
\, , \label{Lohdear}
\end{align}
where $M_{D,U} \gg v_X$ are mass parameters of the VLQs.\footnote{One
    might worry that, since the components $\Psi_R^{U,D}$ have the same quantum numbers as the light right-handed quark fields,  we have neglected
dimension-3 mass terms coupling the light right-handed quark fields to $\Psi_L^U$ or $\Psi_L^D$.
But such terms can be removed by a change of basis in the UV theory which has
no other physical effect. In other words, the fields $\Psi_R^{U,D}$ are identified with the linear combinations that couple via a dimension-3 mass term to $\Psi_L^{U,D}$, and $M_{U,D}$ denotes the mass eigenvalue. The light quark fields are identified with the zero mass eigenstates, which only acquire their mass after electroweak symmetry breaking.
} Without loss of generality, we can take the coefficients $\kappa_D$ and $\kappa_U$ to be real, while the $\lambda_{D,U}^i$ couplings are all complex. 

\begin{figure}[ht]
\begin{center}
\begin{tikzpicture}
\begin{feynman}
\vertex (a) { $\overline{Q^\prime_i}$};
\vertex [right=0.8in of a] (b);
\vertex [right=0.6in of b] (c);
\vertex [right=0.6in of c] (d);
\vertex [right=0.6in of d] (e) { $u^\prime_{3 R}$};
\node at (b) [circle,fill,inner sep=1.5pt,label=below:{ $\,\,\,\, \lambda^i_U$}]{};
\node at (d) [circle,fill,inner sep=1.5pt,label=below:{ $\kappa_U\,\,$}]{};
\node at (c) [square dot,fill,inner sep=1.0pt]{};
\vertex [above=0.7in of b] (f) { $\tilde{H}$};
\vertex [above=0.7in of d] (g) { $\theta$};
\diagram* {
(a) -- [fermion] (b)  -- [fermion,edge label={ $\Psi^U_R$}] (c) -- [anti fermion,edge label={ $\overline{\Psi^U_L}$}] (d) -- [anti fermion] (e),
(b) -- [scalar] (f),
(d) -- [scalar] (g),
};
\end{feynman}
\end{tikzpicture}
\end{center}
\caption{Feynman diagram that gives the dimension-5 effective up-type Yukawa
  operators in (\ref{eq:dim5yuk}), which themselves match onto the 1-3 and
  2-3 up-type Yukawa couplings. The corresponding diagrams for down quark
  Yukawa operators can be obtained by trading every sub/superscript `$U$' for `$D$', and swapping $\tilde H$ for $H$. \label{fig:dim5yuk} }    
\end{figure}
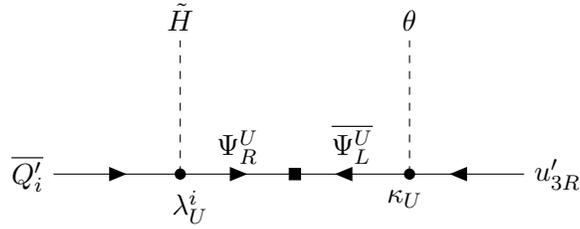

Integrating out the VLQs $\Psi^D$ and $\Psi^U$ at tree level, at scales $\Lambda_D := M_D$ and $\Lambda_U := M_U$ respectively, we
generate the dimension-5 operators written in (\ref{eq:dim5yuk}) from
Feynman diagrams such as the one depicted in Fig.~\ref{fig:dim5yuk}. The (dimensionless) WCs are
\begin{equation}
C^i_U = \lambda^i_U \kappa_U, \qquad C^i_D = \lambda^i_D \kappa_D, \qquad \tilde{C}^i_U=\tilde{C}^i_D = 0\, \qquad i \in \{1,2\}\, .
\end{equation} 
Once $U(1)_{X}$ is broken by $\theta$ acquiring its vacuum expectation value, in either of our benchmark models, these operators match onto dimension-4 up-type and down-type Yukawa couplings suppressed by one power of $\epsilon_U := \kappa_U v_X/(\sqrt{2}M_U)$ or $\epsilon_D:=\kappa_D v_X/(\sqrt{2}M_D)$ respectively, thus populating the third column of both quark Yukawa matrices (but not the third row) with small couplings,
\begin{equation}
Y_u \sim \begin{pmatrix}
\times&\times&\epsilon_U\lambda_U^1\\
\times&\times&\epsilon_U\lambda_U^2\\
0&0&\times\\
\end{pmatrix}, \qquad
Y_d \sim \begin{pmatrix}
\times&\times&\epsilon_D\lambda_D^1\\
\times&\times&\epsilon_D\lambda_D^2\\
0&0&\times\\
\end{pmatrix}\, ,
\end{equation}
where each `$\times$' denotes a dimension-4 renormalisable Yukawa coupling. 
It is therefore natural that the CKM angles mixing the first two families with 
the third are small.\footnote{Of course, such a model sheds no light on
  the \emph{mass} hierarchies of either up-type or down-type quarks.}

In particular, the left-handed mixing angle between the second and third
    family down-type quarks is
  \begin{equation}
\sin \theta_{23} \approx \kappa_D \lambda_D^2 \frac{v_X}{M_D}\, ,
\end{equation}
while the corresponding angle for up-type quarks
is $\kappa_U \lambda_U^2 v_X/M_U$.  
Working perturbatively in the small angles,
the CKM angle $V_{cb}$ is then
\begin{equation} \label{eq:Vcb}
V_{cb} \approx v_X \left(\frac{\kappa_D \lambda_D^2}{M_D} - \frac{\kappa_U \lambda_U^2}{M_U} \right)\, .
\end{equation}
(Note that this admits a complex phase because the Yukawa couplings $\lambda_{U,D}^i$ are complex).  
Thus, if there is a mild hierarchy between $M_D$ and $M_U$ or between
$\lambda_D^2$ and $\lambda_U^2$ then $\sin \theta_{23}$ --- which enters the
$B$-anomaly 
phenomenology of our models --- is predicted to be an order of magnitude or so
smaller than $|V_{cb}|$.

We can invert this argument to estimate the rough mass
  scale of the VLQs. If we assume that the 
  couplings $\lambda_{U,D}^{1,2}$ are order unity, then we expect the
  lighter 
  of the two VLQ masses to be of order 
$M \sim v_X/(\sqrt{2} |V_{cb}|) \approx 18v_X$. In the case of the $\ZP$
  model, we can relate this to the mediator mass via
    (\ref{eq:MZp}):
$M \sim 18 M_{\ZP}/g_X \gg M_{\ZP}$. Indeed, if one looks ahead to the
  global fits (see Fig.~\ref{fig:param}), we see that
  $M_{Z^\prime}/g_{Z^\prime}$ is larger  than
  5~TeV in the 95$\%$ CL fit region, meaning that we expect both VLQ
    masses to be
\begin{equation}
M_{U,D} \gtrsim 90 \text{~TeV}\, .
\end{equation}
The VLQs in these simple models
  decouple from the low-energy phenomenology.
  (For the leptoquark
  model, the scale $v_X$ of $U(1)_X$ breaking
  is not tied to the \bsll\ phenomenology at all
  and so $v_X$, and thus $M_{D,U}$ can be higher still.)

\section{Tree-level matching to the SMEFT \label{sec:SMEFT}}

While recent work has derived one-loop SMEFT matching formulae of models such
as the ones we consider~\cite{Gherardi:2020det,Carmona:2021xtq}, for
our purposes tree-level matching~\cite{deBlas:2017xtg} is a sufficiently
accurate approximation.\footnote{One potentially significant effect that we have
  considered is from the one-loop contributions to four-quark operators in the
  leptoquark model, which are absent at tree-level but which give the leading
  contributions to $B_s-\overline{B}_s$ mixing.
  The relevant WCs generated by one-loop diagrams are~\cite{Gherardi:2020det,Gherardi:2020qhc,Kosnik:2021wyp}, for general flavours,
$(C_{qq}^{(1)})^{ijkl} =9 (C_{qq}^{(3)})^{ijkl} = -\frac{1}{16\pi^2}\frac{9\lambda^4}{16}{(\Lambda_\Xi^{(d_L)})}_{il}{(\Lambda_\Xi^{(d_L)})}_{kj}\,$.
Specialising to those relevant to $B_s-\overline{B}_s$ mixing,
\begin{align} \label{eq:Bs-WC-LQmodel}
(C_{qq}^{(1)})^{2323} = 9 (C_{qq}^{(3)})^{2323}  =  -\frac{1}{16\pi^2}\frac{9\lambda^4}{16}{(\Lambda_\Xi^{(d_L)})}_{23}{(\Lambda_\Xi^{(d_L)})}_{23}\,.
\end{align}
We include these one-loop contributions in our analysis. \label{foot:1loop}
}

In Table~\ref{tab:S3_wcs} we record the tree-level dimension-6 SMEFT
coefficients for our benchmark 
scalar leptoquark model resulting from integrating the $S_3$ field out of the theory, as computed in Ref.~\cite{Gherardi:2020det}. 
In Table~\ref{tab:B3L2_wcs} we record the tree-level dimension-6 SMEFT
coefficients for our benchmark \BthreeLtwo\ obtained by integrating out the
$Z^\prime$ boson. For both the leptoquark and the $Z^\prime$
    model, the VLQs introduced in \S\ref{sec:mixing} to
    account for the CKM mixing with the third family  
are significantly more massive than the leptoquark or $Z^\prime$, with masses of at least 90 TeV as discussed above. Their effects on the SMEFT matching are therefore sub-leading. We take care to
check the size of their effect in $B_s$ meson mixing in \S\ref{sec:Bs-mix-LO}, 
which we find to be negligible.

\begin{table}[ht]
\renewcommand{\arraystretch}{1.3}  
   \begin{center}
     \begin{tabular}{|c|c|} \hline
       WC & value \\ \hline
       ${\bm (C_{lq}^{(1)})^{22ij}}$ & $\frac{3}{4} {\Lambda_\Xi^{(d_L)}}_{ij}$ \\
         $(C_{lq}^{(3)})^{22ij}$ & $\frac{1}{4} {\Lambda_\Xi^{(d_L)}}_{ij}$ \\
 \hline  \end{tabular}
   \end{center}
   \caption{\label{tab:S3_wcs}
     Non-zero tree-level dimension-6 SMEFT WCs predicted by integrating out the $S_3$
     in our benchmark model (\ref{eq:S3lag}), in units of $\lambda^2$, in the Warsaw
     basis~\cite{Grzadkowski:2010es}. Here the EFT matching scale is $\Lambda=M_{S_3}$.
}
\end{table}
\renewcommand{\arraystretch}{1}  
\begin{table}[ht]
\renewcommand{\arraystretch}{1.3}  
   \begin{center}
     \begin{tabular}{|c|c||c|c|} \hline
       WC & value & WC & value \\ \hline
       $C_{ll}^{2222}$&$-\frac{9}{2}$        &
       $ { (C_{lq}^{(1)})}^{22ij}$&$ 3 {\Lambda_{\Xi\ ij}^{(d_L)}}$ \\
       $(C_{qq}^{(1)})^{ijkl}$&$ {\Lambda_{\Xi\ ij}^{(d_L)}} {\Lambda_{\Xi\ kl}^{(d_L)}}\frac{\delta_{ik}\delta_{jl}-2}{2}$ &
       $C_{ee}^{2222}$&$-\frac{9}{2}$ \\     
       $C_{uu}^{3333}$&$-\frac{1}{2}$ &     
       $C_{dd}^{3333}$&$-\frac{1}{2}$ \\     
       $C_{eu}^{2233}$&$3$ &     
       $C_{ed}^{2233}$&$3$ \\     
       $(C_{ud}^{(1)})^{3333}$&$-1$ &
       ${C_{le}^{2222}}$&$-9$ \\     
       $C_{lu}^{2233}$&$3$ &     
       $C_{ld}^{2233}$&$3$ \\     
       $C_{qe}^{ij22}$&$3{\Lambda_{\Xi\ ij}^{(d_L)}}$ &     
       $(C_{qu}^{(1)})^{ij33}$&$- {\Lambda_{\Xi\ ij}^{(d_L)}}$ \\     
       $(C_{qd}^{(1)})^{ij33}$&$-  {\Lambda_{\Xi\ ij}^{(d_L)}}$ &
        & \\
 \hline  \end{tabular}
   \end{center}
   \caption{\label{tab:B3L2_wcs}
     Non-zero tree-level dimension-6 SMEFT WCs predicted by integrating the $Z^\prime$
     out of the \BthreeLtwo, in units of      $g_{Z^\prime}^2$, in the Warsaw
     basis~\cite{Grzadkowski:2010es}. Here the EFT matching scale is $\Lambda=M_{\ZP}$.
}
\end{table}
\renewcommand{\arraystretch}{1}  


  In order to make contact with the WCs of the WET, we must first match 
  the $B_3-L_2$ model to the 
  SMEFT WCs. Strictly
speaking, this should be done at 
the mass of the $\ZP$ or the $S_3$ leptoquark. One then renormalises down to
the $Z^0$ boson mass before matching to the WET\@.
While we shall ignore such renormalisation effects in our
analytic discussion (because they are small), one-loop renormalisation effects
are fully taken into account 
in our numerical implementation of both models and thus in our fit results.'
The most relevant WET operators for our discussion are those in the
  $bs\mu^+\mu^-$ system,
including the operators $\mathcal{O}_9$ and $\mathcal{O}_{10}$ defined in Eqs.~\ref{eq:WET}--\ref{o9and10}, but also
 the
scalar operator $\mathcal{O}_S$ and pseudo-scalar operator $\mathcal{O}_P$
\begin{align}
  \mathcal{O}_S = \frac{e^2}{16 \pi^2} \left(\bar s P_R b\right)
  \left( \bar \mu \mu \right), 
\qquad
  \mathcal{O}_{P} = \frac{e^2}{16 \pi^2} \left(\bar s P_R b\right)
  \left( \bar \mu \gamma_5 \mu \right), \label{oSandP}
\end{align}
and the operators in (\ref{o9and10}), (\ref{oSandP})
with \emph{primes}
(i.e.\ $\mathcal{O}_9^\prime, \mathcal{O}_{10}^\prime, \mathcal{O}_S^\prime,
\mathcal{O}_P^\prime$) where 
one makes the replacement $P_L \leftrightarrow P_R$.

The tree-level SMEFT-to-WET coefficient matching formulae are
well
known~\cite{Aebischer:2015fzz} 
\begin{align}
  C_9 &= K\left(C_{qe}^{2322} + (C_{lq}^{(1)})^{2223} + (C_{lq}^{(3)})^{2223}+
  (1-4s_W^2)[(C_{Hq}^{(1)})^{23} + (C_{Hq}^{(3)})^{23}]\right), \nonumber \\
   C_{10} &= K\left(C_{qe}^{2322} - (C_{lq}^{(1)})^{2223} - (C_{lq}^{(3)})^{2223}+
   (C_{Hq}^{(1)})^{23} + (C_{Hq}^{(3)})^{23}\right),  \nonumber \\
   C_9^\prime &= K\left(C_{ed}^{2223} + C_{ld}^{2223} -
   (1-4s_W^2)C_{Hd}^{23}\right), \nonumber \\
   C_{10}^\prime &= K\left(C_{ed}^{2223} - C_{ld}^{2223}+C_{Hd}^{23}\right), \nonumber \\
   C_S&=-C_P=K C_{ledq}^{2232}, \nonumber \\
   C_S^\prime&=C_P^\prime=K C_{ledq}^{2223},  \label{smeftWET}
\end{align}
where $s_W=\sin \theta_W$ with $\theta_W$ the Weinberg angle, and
$K = 2\sqrt{2}\pi^2/(e^2 G_F \Lambda^2)$ where $\Lambda$ is the EFT matching scale which is identified with the heavy
particle mass in either case.

In our conventions, $C_9 > 0$ can fit the flavour
data 
well~\cite{Aebischer:2019mlg}\footnote{Note that in some conventions a factor
  of $(V_{ts} V_{tb}^\ast)$ is included in the definition of $C_9$, 
as in Ref.~\cite{Aebischer:2019mlg}, which reverses the
sign of $\mathrm{Re}(C_9)$. }. 
Substituting in the SMEFT WCs from Table~\ref{tab:S3_wcs}, we see that, for 
the $S_3$ leptoquark model 
\begin{equation}
  C_9 = -C_{10} =  K \lambda^2 \Lambda^{(d_L)}_{\Xi_{23}}= -K
 \lambda^2 \frac{\sin 2 \theta_{23}}{2},\qquad C_9^\prime =
  C_{10}^\prime=C_S=C_P=C_S^\prime=C_P^\prime = 0. \label{LQpred}
\end{equation}
On the other hand, for the \BthreeLtwo, substituting the SMEFT WCs in from
Table~\ref{tab:B3L2_wcs} and using (\ref{angle}),
\begin{align}
  C_9 = K 6 g_{Z^\prime}^2 \Lambda_{\Xi_{23}}^{(d_L)} =
  - K 3 g_{Z^\prime}^2 \sin 2 \theta_{23},
  \
  C_{10} = C_9^\prime =
  C_{10}^\prime=C_S=C_P=C_S^\prime=C_P^\prime = 0. \label{ZPpred}
\end{align}

A crucial difference between the WCs of the two benchmark models
is that in
the $\ZP$ model one turns on not only 
semi-leptonic operators, but also four-quark and four-lepton operators,
whereas a leptoquark gives only semi-leptonic operators in isolation. Since
four-quark operators can give contributions to processes such as
$B_s-\overline{B_s}$ meson mixing (via the process depicted in the right-hand panel
of Fig.~\ref{fig:BsBsbar}) which do not exhibit a significant disagreement
between measurements and SM predictions, it
is often stated that leptoquark models 
provide a `better explanation' of the data. One purpose of the present paper
is to 
quantitatively assess this claim, by comparing the statistical preference of
the \bsll\ data including $B_s-\overline{B_s}$ mixing for either model. 
Before describing the fits in detail, we
  shall review the dependence of certain key observables,  
  namely $B_s-\overline{B}_s$ mixing and the $B_s\rightarrow \mu^+\mu^-$
  branching ratio, on the WCs. These are important observables that, 
  given sufficient precision, could discriminate between
  the $B_3-L_2$ $\ZP$ model and the $S_3$ leptoquark model.

\subsection{Leading contributions to $B_s - {\overline B_s}$ mixing} \label{sec:Bs-mix-LO}

One contribution to $B_s-{\overline B_s}$ mixing stems from
    the mass difference of the mass 
    eigenstates of the neutral $B_s$ mesons, $\Delta M_s$. 
A new physics model contributes to $\Delta M_s$ if it generates an effective Lagrangian proportional to $(\bar s \gamma_\mu P_X b) (\bar s
  \gamma^\mu P_Y b)$ plus the Hermitian conjugate term, where $\{X,Y\} \in
  \{L,R\}$.
  Compared to the SM prediction, the 4-quark SMEFT operators change the
  prediction for $\Delta M_s$ 
  by a 
factor (derived from Refs.~\cite{Aebischer:2015fzz,DiLuzio:2017fdq})
\begin{equation}
  \frac{\Delta M_s}{\Delta M_s^{(\mathrm{SM})}} = \left|1 -
  \frac{\eta(\Lambda)[(C_{qq}^{(1)})^{2323}+(C_{qq}^{(3)})^{2323}]}{R^\text{loop}_{{\mathrm{SM}}} \Lambda^2 }
  \frac{\sqrt{2}}{4G_F 
    (V_{tb} V_{ts}^*)^2} \right|, \label{eqone}
\end{equation}
where $R_{\mathrm{SM}}^{\text{loop}} = 1.3397 \times 10^{-3}$
and $\eta(\Lambda)$
parameterises renormalisation effects
between the scale $\Lambda$ where the SMEFT WCs are set and the
bottom quark mass. 


  The \BthreeLtwo\  induces
  $(C_{qq}^{(1)})^{2323}$ at 
  tree-level (see Table~\ref{tab:B3L2_wcs}), where the cut-off scale $\Lambda$ is identified with $M_{\ZP}$.
$\eta(M_{Z^\prime})$ varies between 0.79 and 0.74 when
$M_{Z^\prime}$ ranges between 1 and 10 TeV~\cite{DiLuzio:2017fdq}.
Substituting in for $(C_{qq}^{(1,3)})^{2323}$ from Table~\ref{tab:B3L2_wcs} and (\ref{angle}),
we find that
\begin{equation}
  \frac{\Delta M_s}{\Delta M_s^{(\mathrm{SM})}} = \left|1 +
  \frac{\eta(M_{Z^\prime}) g_{Z^\prime}^2 \sin^2 2 \theta_{23}}
    {8 R^\text{loop}_{\mathrm{SM}} M_{Z^\prime}^2 }
  \frac{\sqrt{2}}{4 G_F 
    (V_{tb} V_{ts}^*)^2} \right|. \label{eqtwo}
\end{equation}
The SM prediction of $\Delta M_s$ is 
higher than the average of experimental measurements (but roughly agrees with
it) and so 
$\Delta M_s$
restricts the size of $g_{Z^\prime}^2 \sin^2 2 \theta_{23} /
M_{Z^\prime}^2$ to not be too large~\cite{DiLuzio:2017fdq}.

Finally, we note that in both models there are one-loop contributions to
$B_s-\overline{B}_s$ mixing coming 
from box diagrams involving exchange of the VLQs. 
These contributions scale as $1/(16 \pi^2 M_{U,D}^2)$, with further suppression factors coming from suppressed flavour-changing interactions,
and so are smaller still than the one-loop contribution coming from integrating out the $S_3$ leptoquark. Indeed, the largest of these contributions comes from 
a diagram involving $W$-boson exchange and the $\Psi^U_{L,R}$ fields running in the loop, which scales as
$\frac{\sin^2 \theta_{23}}{16 \pi^2 M_U^2}(1 + \mathcal{O}(g^2))$, where $g$ is the $SU(2)_L$ gauge coupling. We therefore drop these tiny contributions to 4-quark operators 
in the rest of this paper.

As discussed in footnote~\ref{foot:1loop}, the $S_3$ model
  does not generate $(C_{qq}^{(1)})^{2323}$ or $(C_{qq}^{(3))})^{2323}$  at
  tree-level; the dominant contribution appears at one loop as in
  (\ref{eq:Bs-WC-LQmodel}). 
 Substituting this into (\ref{eqone}), we obtain
  \begin{equation}
  \frac{\Delta M_s}{\Delta M_s^{(\mathrm{SM})}} = \left|1 +
\frac{\lambda^4}{M_{S_3}^2}
  \frac{5\eta(M_{3}) \lambda^4 \sin^4 2 \theta_{23}}
    {8192 \pi^2\,  R^\text{loop}_{\mathrm{SM}} M_{3}^2}
  \frac{\sqrt{2}}{4 G_F 
    (V_{tb} V_{ts}^*)^2} \right|. \label{eqthree}    
    \end{equation}
  As expected, the non-SM term in
  (\ref{eqthree}) is suppressed relative to the equivalent term in 
the $\ZP$ model by a factor of $[\text{coupling}]^2/(16\pi^2)$. Numerically,
we ran the global fits (see \S\ref{sec:fits}) both with and without these
1-loop contributions for the $S_3$ model, and found the differences in the fit
quality (and in particular, in the fit to the $\Delta m_s$ observable) to be
negligible for $M_{S_3}=3$~TeV, the canonical value that we shall take in
\S\ref{sec:fits}. They could become important, however, for much larger values of $M_{S_3}$,
since the rest of the observables in the fit prefer $\lambda$ proportional to
$M$.

\subsection{New physics effects in $BR(B_s \rightarrow \mu^+ \mu^-)$ \label{sec:Bstomumu}}
Below, we shall update the \smelli\ prediction of the branching ratio of $B_s$ and $B_d$ mesons
decaying to muon/anti-muon pairs with more recent data. 
In the presence of new physics parameterised in terms of WET WCs,
the $B_s\rightarrow \mu^+\mu^-$ branching ratio prediction is changed from the SM one
by a multiplicative factor\footnote{Here, we are listing the \emph{prompt} decay branching
    ratios (i.e.\ without the bar) for clarity of the analytic discussion. For
    the relation between the prompt branching ratio
    and the CP-averaged time-integrated branching ratio, see
Ref.~\cite{DeBruyn:2012wk}.}~\cite{Altmannshofer:2017wqy}
\begin{align}
  \frac{BR(B_s \rightarrow \mu^+\mu^-)}{BR(B_s \rightarrow \mu^+\mu^-)_{\text{SM}}}
  &=  \left|  1 + \frac{C_{10}-C_{10}^\prime}{C_{10}^{\mathrm{SM}}} + \frac{C_{P}-C_P^\prime}{C_{10}^{\mathrm{SM}}}\frac{m_{B_s}^2}{2m_\mu(m_b+m_s)}  \right|^2  \nonumber\\
& \quad + \left(1-\frac{4m_\mu^2}{m_{B_s}^2} \right) \left|
\frac{C_{S}-C_S^\prime}{C_{10}^{\mathrm{SM}}}\frac{m_{B_s}^2}{2m_\mu(m_b+m_s)}  \right|^2
\, , \label{bstomm}
\end{align}
where $m_{B_s}$, $m_b$, $m_s$ and $m_\mu$ are the masses of the $B_s$ meson,
the 
bottom quark, the strange quark and the muon, respectively.
Eqs. (\ref{LQpred}) and (\ref{ZPpred}) imply that
$C_P$, $C_S$, $C_P^\prime$, $C_S^\prime$,
  $C_9^\prime$ and $C_{10}^\prime$ are all negligible both in
  the \BthreeLtwo\ and in our $S_3$ model. 
  Thus, (\ref{bstomm}) implies that, in these two models,
  \begin{equation}
  BR(B_s \rightarrow \mu^+\mu^-)= \left|  1 + \frac{C_{10}}{C_{10}^{SM}} \right|^2
  BR(B_s \rightarrow \mu^+\mu^-)_{{SM}}  \, .  \label{bstomm2}
\end{equation}
  Since (aside from small loop-level corrections induced by renormalisation
  group running) $C_{10}=0$ in the
  \BthreeLtwo, $BR(B_s \rightarrow \mu^+\mu^-)=BR(B_s \rightarrow \mu^+\mu^-)_{SM}$.
  On the other hand, $C_{10} \neq 0$ in our $S_3$ leptoquark model and so
  $BR(B_s \rightarrow \mu^+\mu^-)$ can significantly deviate from its SM
  limit. This difference between the models, which has a noticeable impact on the fits (see \S\ref{sec:fits}), is just a consequence of 
fitting the anomalies with a pure $C_9$ new physics contribution in the $\ZP$
model verses
a $C_9 = -C_{10}$ new physics effect in the $S_3$ leptoquark model.

\section{Fits \label{sec:fits}}

We now turn to fits of each benchmark model to 
flavour transition data. Each simple benchmark model
has been characterised by three parameters: $\theta_{23}$, the overall strength of the
coupling of the new physics state to fermions ($\lambda$ in the $S_3$ model
and $g_{Z^\prime}$ in the $\ZP$ model, respectively) as well as the
mass of the new physics field ($M_{S_3}$ in the leptoquark model and $M_{Z^\prime}$ in
the $\ZP$ model). Having encoded the SMEFT WCs as a function of these parameters, we
feed them into \smelli\ at a renormalisation scale corresponding to the mass of
the new physics state. The \smelli\ program then renormalises the WCs, matches to the
WET, and calculates observables at the
bottom quark mass. It then returns
a $\chi^2$ score to
characterise the quality of fit. To a good approximation, the fit results are
independent of the mass of the new physics state, provided one re-scales the
overall strength of the coupling of the new physics state to fermions in each
case linearly, proportional to the mass. The dominant missing relative
corrections to the WCs from 
this approximation are small: for two different $\ZP$ masses $M_{\ZP}^{(1)}$ and $M_{\ZP}^{(2)}$, say,
the missing correction is of order $1/(16 \pi^2) \log (M_{\ZP}^{(2)}/M_{\ZP}^{(1)})$.
To this good approximation then, each fit is over two effective
parameters: one is $\theta_{23}$ and the other is
$g_{Z^\prime}/M_{Z^\prime}$ or $\lambda/M_{S_3}$, depending upon the model.\footnote{We note from (\ref{eq:Bs-WC-LQmodel}), however, that the one-loop corrections to $B_s-\overline{B_s}$
  mixing instead scale differently, as $\lambda^4/M_{S_3}^2$. These corrections are negligible
  unless $M_{S_3}$ is very large $\sim \mathcal{O}(100\text{~TeV})$, since
  the \bsll anomalies prefer $\lambda \propto M_{S_3}$.}

Direct search limits are
a different function of the
coupling and mass
of the new physics state, however. 
Currently, the most stringent 95$\%$ CL LHC experimental lower limits from
the production of
di-leptoquarks (which each decay to a jet and (anti-)muons or (anti-)neutrinos)
is around 1.4~TeV from the ATLAS collaboration~\cite{ATLAS:2020dsk}. 
The \BthreeLtwo\ has a lower $M_{Z^\prime}$ limit of around 1~TeV 
within the 95$\%$ CL parameter space region of a previous \bsll\ anomaly
fit~\cite{Allanach:2021gmj}.      
Here, we shall illustrate with a reference mass of
3
TeV for either $M_{\ZP}$ or $M_{S_3}$ depending on the model, noting that such
a mass is allowed by direct searches. We comment further on the constraints coming from the LHC in \S\ref{sec:highpt}.

We use the default \smelli\ development version constraints upon all
observables except for $R_{K^{(\ast)}}$ and 
$\{BR(B\rightarrow \mu^+\mu^-), \overline{BR}(B_s \rightarrow \mu^+ \mu^-)\}$,
which have 
a new joint experimental measurement. We now detail our implementation of the
new 
measurement (we also ran {\tt \smelli} to recalculate all of the covariances
in the theoretical uncertainties after this change).

\subsection{Fit to $\overline{BR}(B_s \rightarrow \mu^+\mu^-)$ data\label{sec:bsmm}}

In July 2022,
CMS released an analysis of 140 fb$^{-1}$ of LHC
data~\cite{CMS-PAS-BPH-21-006}, putting constraints 
jointly upon $\overline{BR}(B_s \rightarrow \mu^+ \mu^-)$ and $BR(B \rightarrow \mu^+
\mu^-)$. We combine this with the most recent similar 
measurements from ATLAS~\cite{ATLAS:2018cur} and LHCb~\cite{LHCb:2021vsc}. 
We approximate the two-dimensional likelihood from
each constraint as a Gaussian in order to easily combine them.
Following Ref.~\cite{Altmannshofer:2021qrr}, we approximate each measurement
as being independent, which should be a reasonable approximation because each
measurement's uncertainty is statistically dominated. 
\begin{figure}
  \begin{center}
    \unitlength=\textwidth
      \includegraphics[width=0.5 \textwidth]{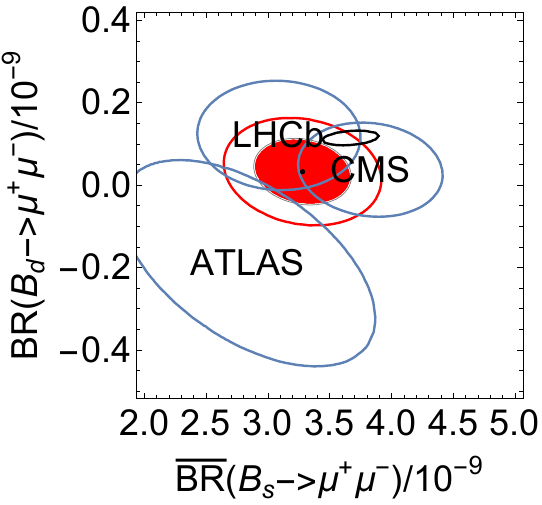}
  \end{center}
  \caption{\label{fig:bsconst} Combination of joint measurements of
    $\overline{BR}(B_s\rightarrow \mu^+\mu^-)$ and $BR(B_d \rightarrow \mu^+ \mu^-)$. The
    blue ellipses show the 68$\%$ CL ellipse from each individual experiment
    (labelled in the centre of each ellipse, respectively),
    whereas the filled red ellipse shows their combination. The empty red
    ellipse shows the 95$\%$ CL contour of the combination.
    The 68$\%$ CL SM prediction with \flavio\ defaults is shown as the small
    black ellipse. 
  }
\end{figure}
The Gaussian approximations to the measurements and our combination of them
are depicted in 
Fig.~\ref{fig:bsconst}. The combination corresponds to
\begin{eqnarray}
  \overline{BR}(B_s \rightarrow \mu^+ \mu^-)&=&(3.28 \pm 0.26) \times 10^{-9}\, , \nonumber
  \\
    BR(B \rightarrow \mu^+ \mu^-)&=&(3.21 \pm 5.34) \times
    10^{-11}\, , \label{bsComb} 
\end{eqnarray}
with a correlation coefficient of $\rho=-0.16$. 
Following Ref.~\cite{Altmannshofer:2021qrr} again and using default
\flavio\ settings, one obtains SM predictions of 
\begin{eqnarray}
  \overline{BR}(B_s \rightarrow \mu^+ \mu^-)_{SM}&=&(3.67 \pm 0.15) \times 10^{-9}\, , \nonumber
  \\
    BR(B \rightarrow \mu^+ \mu^-)_{SM}&=&(1.14 \pm 0.12) \times
    10^{-10}\, , \label{bsSM} 
\end{eqnarray}
with a correlation coefficient $\rho=+0.28$. These are jointly displayed in
Fig.~\ref{fig:bsconst} as a small black ellipse.
Taking the experimental uncertainties into account and 
comparing the SM predictions with
the two dimensional 
experimental likelihood,
we obtain a one-dimensional pull of 
1.6$\sigma$, if both   
branching ratios are SM-like. The recent update of the CMS measurement (which
analysed 
significantly more integrated luminosity than it had previously) has reduced
this one-dimensional pull from
2.3$\sigma$~\cite{Altmannshofer:2021qrr}\footnote{In the global fit however,
  one should remember that there are additional correlations with other
  observables and so this change does not straightforwardly translate to
  an identical reduction of $\chi^2$.}.


\subsection{Fit results} \label{sec:fit_results}
We display the best-fit points together with the associated $\chi^2$ breakdown
and associated 
$p-$value in Table~\ref{tab:compare}. From the table, we note that
both models can provide a reasonable fit to the data with $p-$values greater
than {}.1,
in contrast to the SM (see Table~\ref{tab:SMpvals}). We also notice
that the improvement of the fit to data over that of 
the SM is considerable and similar in each case: $\sqrt{\Delta \chi^2}=\Sthreedchisq{}$ for the $S_3$ model and
$\sqrt{\Delta \chi^2}=\BthreeLtwodchisq{}$ for the $Z^\prime$ model.
We see that while the December 2022 LHCb reanalysis of $R_{K^{(\ast})}$
  has reduced the improvement of each model with respect to the SM (evidenced
  by a lower
  $\sqrt{\Delta \chi^2}$), the $p-$value, and therefore the actual quality of the fit, has improved. This is because the SM was a poor fit previously and now it fits the `LFU' observables well (see Table~\ref{tab:SMpvals}) and because, previously, the new physics models were not very well equipped to explain some of the LFU data (in particular, $R_{K^\ast}$ in the low $Q^2$ bin) which showed a larger deviation than expected in each model.

\begin{table}
  \begin{center}
    \begin{tabular}{cc}
      \multicolumn{2}{c}{Dec 2022 $R_{K^{(\ast)}}$~\cite{LHCb:2022qnv}}\\
      \begin{tabular}{|c|cccc|} \hline
$S_3$ model& $\chi^2$ & $n$ & $p$ & $s\sqrt{|\Delta \chi^2|}$ \\ \hline
      quarks & 247.3 & 224 & .14 & 3.9  \\
      LFU & 19.7 & 23 & .66 & -1.6  \\
      global & 267.0 & 247 & .16 & 3.6  \\
      \hline
    \end{tabular}  &     \begin{tabular}{|c|cccc|} \hline
      $Z^\prime$ model& $\chi^2$ & $n$ & $p$ & $s\sqrt{|\Delta \chi^2|}$ \\ \hline      
quarks & 249.1 & 224 & .12 & 3.7  \\
LFU & 18.2 & 23 & .75 & -1.0  \\
global & 267.4 & 247 & .16 & 3.6  \\
      \hline
    \end{tabular} 
 \\
    \end{tabular}
    \begin{tabular}{cc}
      \multicolumn{2}{c}{Previous $R_{K^{(\ast)}}$~\cite{LHCb:2017avl,LHCb:2019hip,LHCb:2021trn}}\\
      \begin{tabular}{|c|cccc|} \hline
$S_3$ model& $\chi^2$ & $n$ & $p$ & $s\sqrt{|\Delta \chi^2|}$ \\ \hline
      quarks & 245.7 & 224 & .15 & 3.7  \\
      LFU & 22.2 & 23 & .51 & 4.2  \\
      global & 267.9 & 247 & .15 & 5.5  \\
      \hline
    \end{tabular} 
 &     \begin{tabular}{|c|cccc|} \hline
      $Z^\prime$ model& $\chi^2$ & $n$ & $p$ & $s\sqrt{|\Delta \chi^2|}$ \\ \hline      
quarks & 249.3 & 224 & .12 & 3.1  \\
LFU & 22.8 & 23 & .47 & 4.1  \\
global & 272.1 & 247 & .11 & 5.1  \\
      \hline
    \end{tabular} 
 \\
    \end{tabular}
      \caption{\label{tab:compare} Quality-of-fit at the best-fit points of the 
        $S_3$ leptoquark model and the $B_3-L_2$ $Z^\prime$ model
        at $M_{Z^\prime}=M_{S_3}=3$~TeV. The first column displays the
        category of observable.
        $n$ is
        the number of measurements in each set.  The $\chi^2$ values for
        each best-fit point are also shown and $\Delta
        \chi^2:=\chi^2_{SM}-\chi^2$ and $s:=\text{sign}(\Delta \chi^2)$.
        The (updated-$R_{K^{(\ast)}}$) best-fit parameters for the
        leptoquark model are $\lambda=\Sthreegzp{}$,
        $\theta_{23}=\Sthreetheta{}$. For the $Z^\prime$ model, they are
        $g_{Z^\prime}=\BthreeLtwogzp$, $\theta_{23}=\BthreeLtwotheta$. Only
        the combined category includes the two fitting parameter reduction in the number
        of degrees of freedom when  calculating $p$.  The results using the
        $R_{K^{(\ast)}}$ LHCb
        measurements~\cite{LHCb:2017avl,LHCb:2019hip,LHCb:2021trn} prior to
        December 2022 (but including the $BR(B_s\rightarrow \mu^+\mu^-)$ update in \S\ref{sec:bsmm}) were performed  
        separately and have different best-fit parameter values to these.}
  \end{center}
\end{table}

\begin{figure}
  \begin{center}
    \includegraphics[width=0.45\columnwidth]{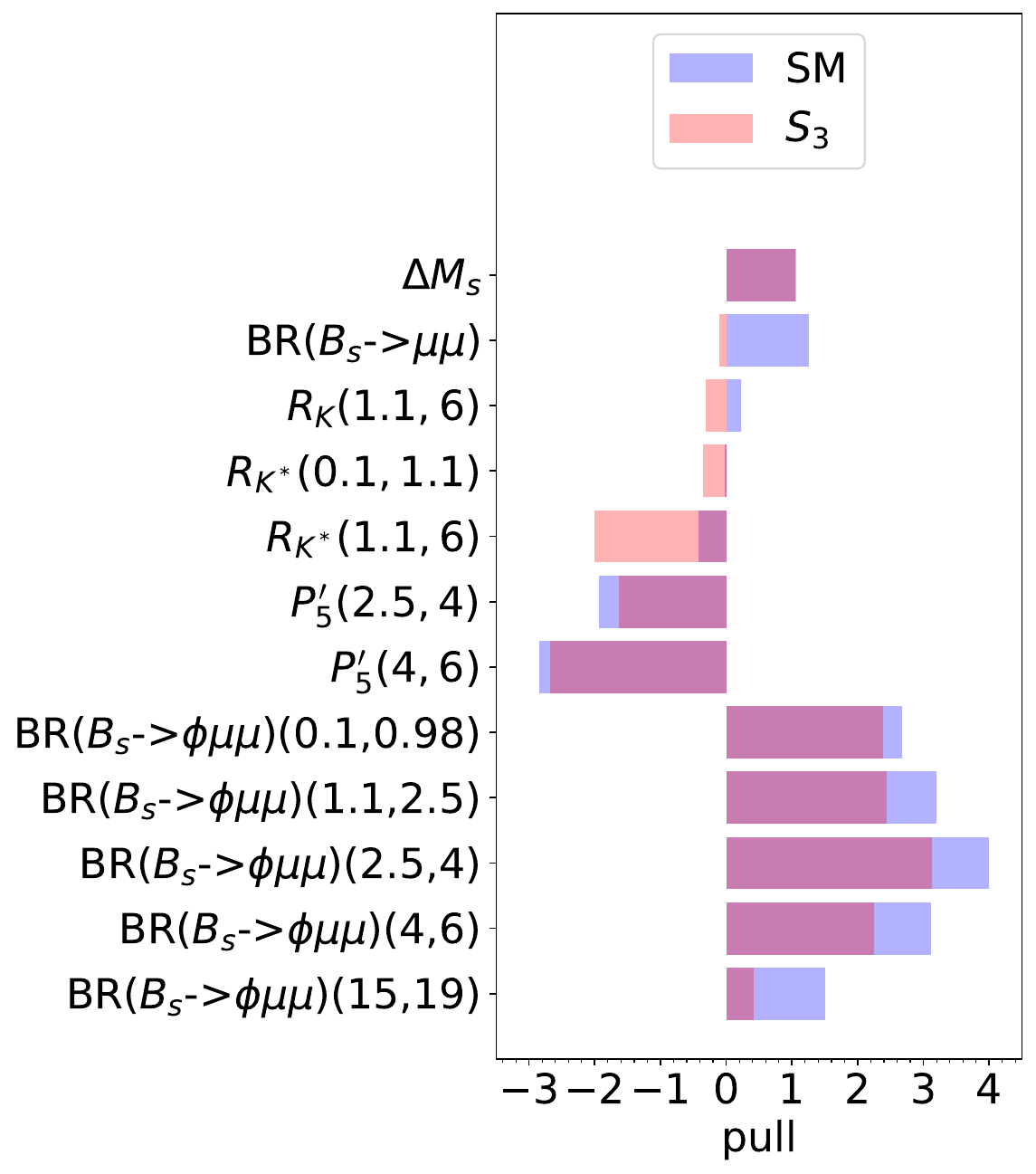}    
    \includegraphics[width=0.45\columnwidth]{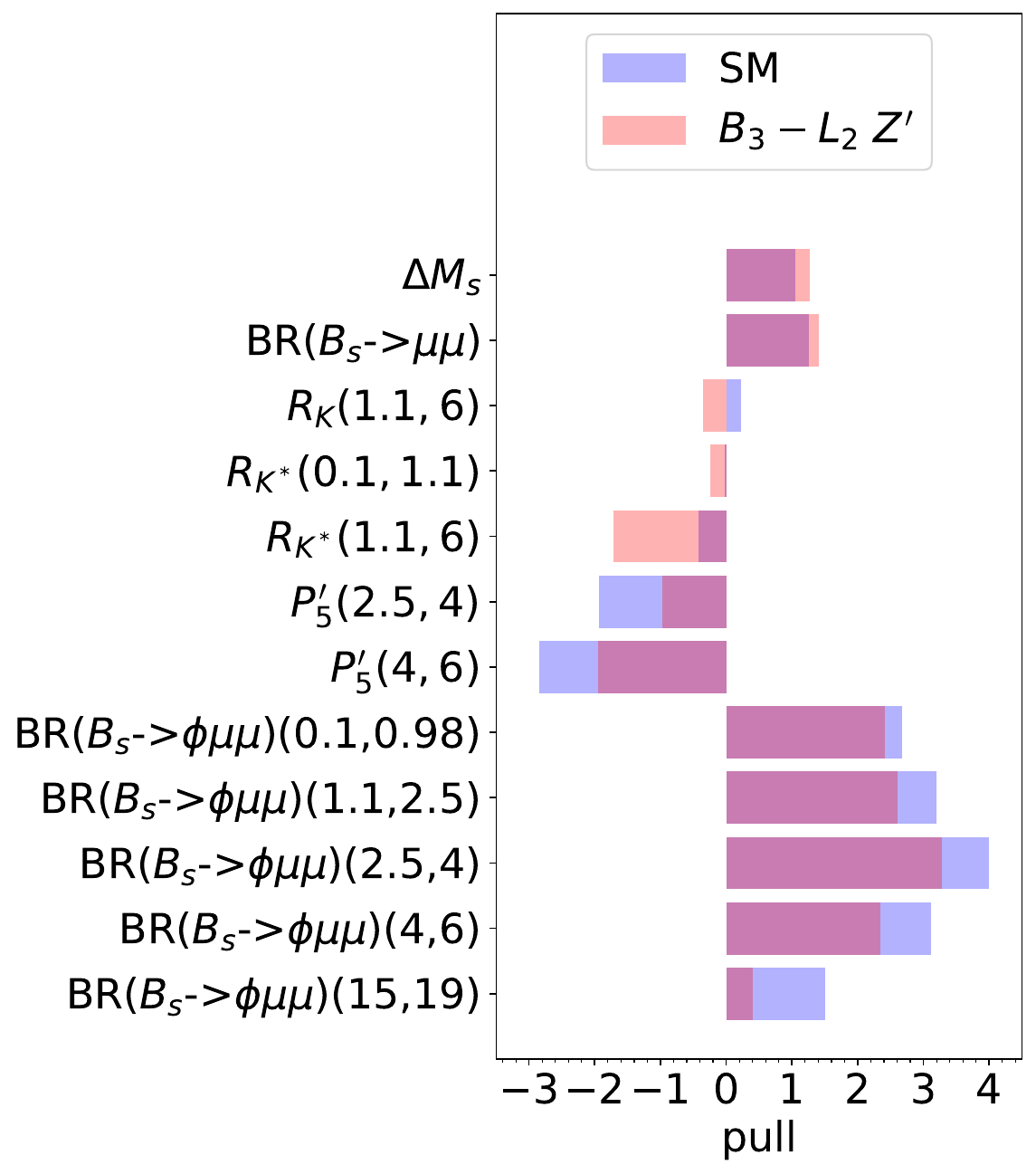}
    \end{center}
  \caption{\label{fig:int} Pulls of some observables of interest for best-fit
    points of the
    $S_3$ leptoquark model at $M_{S_3}=3$~TeV and the 
    \BthreeLtwo{} best-fit point at $M_{Z^\prime}=3$~TeV. The pull is
    defined in (\ref{pull}). Numbers in parenthesis denote the
    domain of the di-lepton invariant mass-squared bin, in GeV$^2$.
            The best-fit parameters for the
        leptoquark model are $\lambda=\Sthreegzp{}$,
        $\theta_{23}=\Sthreetheta{}$. For the $Z^\prime$ model, they are
        $g_{Z^\prime}=\BthreeLtwogzp$, $\theta_{23}=\BthreeLtwotheta$.}
  \end{figure}
We pick out some \bsll\ anomaly observables of interest in
Fig.~\ref{fig:int} to display the pull $P_i$ of observable $i$, defined as
\begin{equation}
  P_i = \frac{T_i - E_i}{S_i} \label{pull}
\end{equation}
where $T_i$ is the theory prediction, $E_i$ is the experimental central value
and $S_i$ is the experimental uncertainty added to the theoretical uncertainty
in quadrature, neglecting correlations with other observables.
From the figure, we notice that $P_{\Delta M_s}$, i.e.\ the pull associated
with $B_s-\overline{B_s}$ mixing, is negligibly far from the SM prediction in
the $S_3$ model best-fit point, as expected since the contributions from the
$S_3$ are at the one-loop level (as are SM contributions) and furthermore are
mass suppressed.  While the $Z^\prime$ contribution to
$\Delta M_s$ is mass suppressed, it is at tree-level and therefore enhanced
compared to the $S_3$ contribution. 
Even though the $Z^\prime$ prediction for $\Delta M_s$ does
noticeably differ from that of the SM, it is only by a
small amount and the effect on quality-of-fit is consequently small.
We see from (\ref{ZPpred}) that 
the new physics contribution to $b\rightarrow s \mu^+\mu^-$
observables is proportional to $g_{Z^\prime}^2 \sin 2
\theta_{23}/M_{Z^\prime}^2$,
whereas 
the new physics contribution to $\Delta M_s$ is proportional to $g_{Z^\prime}^2 \sin^2 2
\theta_{23}/M_{Z^\prime}^2$, as can be seen from
(\ref{eqtwo}), i.e.\ it involves
one additional power of $\sin 2 \theta_{23}$. Thus, in the \BthreeLtwo,
decreasing $\theta_{23}$
but 
increasing $g_{Z^\prime}/M_{Z^\prime}$ can then provide a reasonable fit
both to the  $b\rightarrow s \mu^+\mu^-$ anomalies (which require a sizeable
new physics contribution)
and to $\Delta M_s$, 
which prefers a small new physics contribution.

We also see that the $\overline{BR}(B_s \rightarrow \mu^+\mu^-)$ prediction
agrees with the discussion in
\S\ref{sec:bsmm}, i.e.\ it
is approximately
equal 
to that of the SM in the $Z^\prime$ model, whereas it is quite different (and
slightly ameliorated) in
the $S_3$ model. 
However, the angular distribution
$P_5^\prime$ measurements favour the \BthreeLtwo\ in comparison to the $S_3$
leptoquark 
model. Various observables differ in their pulls between the two
models, but in the final analysis, when tension in one observable is traded
against others, each benchmark model can achieve a similar 
quality-of-fit, as summarised in 
Table~\ref{tab:compare}.

\subsection{Delineating the preferred parameter space regions \label{sec:par}}

\begin{figure}
  \begin{center}
    \unitlength=\textwidth
    \begin{picture}(0.9,0.45)(0,0)
    \put(-0.1,0){\includegraphics[width=0.55 \textwidth]{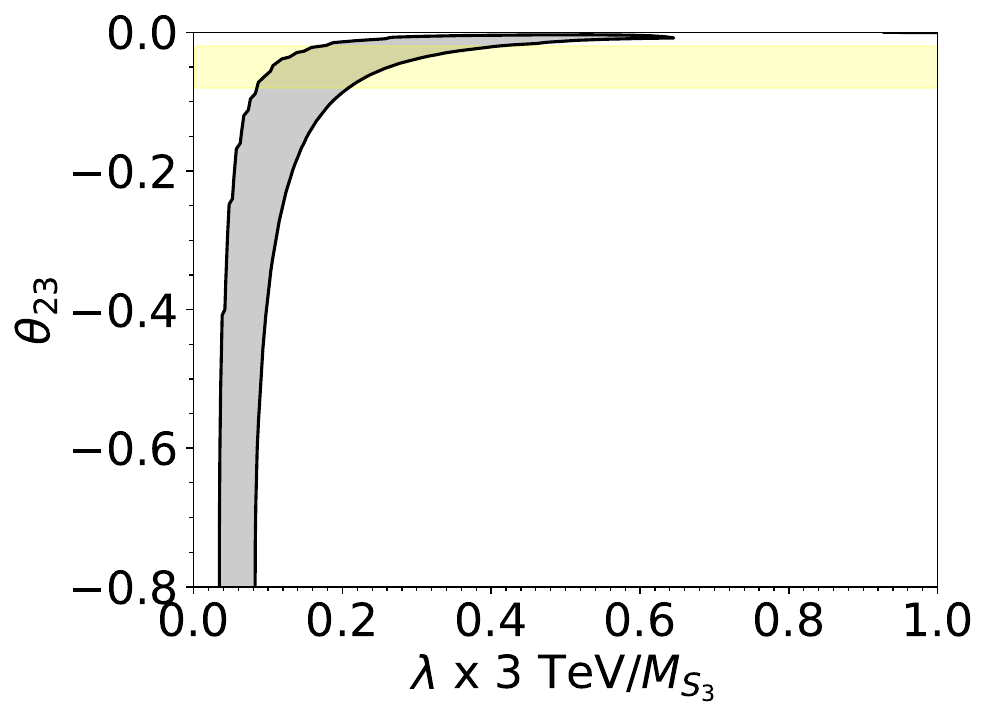}}
    \put(0.45,0.01){\includegraphics[width=0.55 \textwidth]{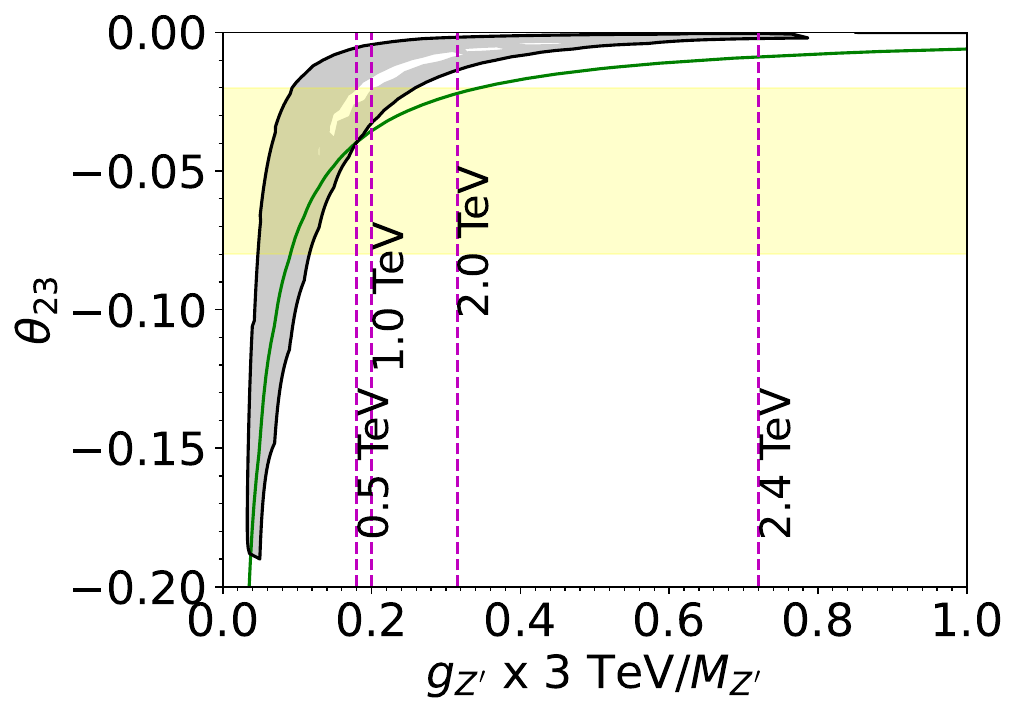}}
    \end{picture}
  \end{center}
  \caption{Two parameter fit to the $S_3$ leptoquark model (left panel) and
    the \BthreeLtwo{} (right panel).
    Shaded
    regions are those preferred by the global fit at the 95$\%$
    confidence level (CL).
      The yellow region displays the domain
        $|\theta_{23}| \in \{|V_{cb}|/2,\ 2|V_{cb}|\}$ to guide the eye.
      The region below the green curve in
      the right-hand plot is incompatible with the $\Delta M_s$ measurement 
      considered on its own at the 95$\%$ CL\@.
    There is no such domain in the leptoquark case, explaining the lack of a
    green curve in the left-hand panel.
    Our datasets are identical (save for $\overline{BR}(B_s \rightarrow
    \mu^+ \mu^-)$ as detailed in \S\ref{sec:bsmm} and $R_{K^{(\ast)}}$) to 
    those defined by \smelli{} 
    and we refer the curious reader to its manual~\cite{Aebischer:2018iyb},
    where the observables are enumerated. The fits have been performed with
    $M_{S_3}=3$ TeV and $M_{Z^\prime}=3$ TeV, respectively; however each fit
    is  
    approximately independent of 
    the precise  value of the new particle mass (see text).
    In the right-hand
    panel, the region to the right-hand side of each vertical dashed magenta
    contour is excluded for $M_{Z^\prime}$/TeV values
    below $x$
    (where $x$ is the labelled value),
    to 95$\%$~CL\@~\cite{Azatov:2022itm}.
    \label{fig:param}}
\end{figure}

We now delineate the preferred regions of parameter space for each benchmark
model, to help facilitate future studies. In Fig.~\ref{fig:param}, we display
a scan over parameter space for each model. The black curves enclose the 95$\%$ CL combined region
of parameter space, defined as $\chi^2 - \chi^2_\text{min}<5.99$, where
$\chi^2_\text{min}$ is the minimum value of $\chi^2$ on the parameter plane.
We see in the figure that in the $S_3$ model, the preferred region reaches
 to large values of $-\theta_{23}$ and particular values of $\lambda / M_{S_3}$. In fact, the $\chi^2$ minimum valley is
 almost flat. 
 On the other hand, $B_s-\overline{B_s}$ mixing limits $-\theta_{23}$ to be
 not too large in the $Z^\prime$ model, producing a less degenerate valley in
 $\chi^2$ (note the different scales between the left-hand and right-hand
   panels' vertical axis).
 We note, however, that values $|\theta_{23}| \sim
 \mathcal{O}(|V_{cb}|) = 0.04$
 are easily within the good-fit region for each model, meaning that there is no need for any
 particular flavour alignment. If $\theta_{23}$ were constrained to be much
 smaller, for example, $V_{cb}$ would be required (speaking from the point of
 view of the initial gauge eigenbasis) to come
 dominantly from  the up-quark sector. For each model, a very thin region of
 allowed parameter space extends to the right-hand side of each plot for small
 values of $-\theta_{23}$ which is not visible (but which can be seen when the
 ordinate is plotted with a logarithmic scaling, for example).


\subsection{High-energy constraints from the LHC}
\label{sec:highpt}

We remind readers of the $M_{S_3}>1.4$~TeV bound resulting from an ATLAS
di-leptoquark search~\cite{ATLAS:2020dsk}. Given that the leptoquark is
coloured, its pair-production proceeds dominantly by QCD production and so
  this bound is insensitive to the coupling $\lambda$  to SM fermions.
There are also high transverse momentum LHC constraints on our scalar leptoquark model
  coming from \emph{e.g.} the Drell--Yan process $pp \to \mu^+\mu^-$, 
which receives a new physics correction due to the $S_3$ leptoquark being
exchanged in the $t$-channel. These Drell--Yan constraints, unlike the
di-leptoquark search, do not scale only with the leptoquark mass but also
scale with the size of its coupling $\lambda$ to quark-lepton pairs, and so
they provide complementary information. 
Using the \texttt{HighPT} package~\cite{Allwicher:2022gkm,Allwicher:2022mcg},
we computed the $\chi$-squared statistic as a function of the
leptoquark model parameters $(\lambda, \theta_{23})$ using the CMS di-muon
search~\cite{CMS:2021ctt} implemented within, for a benchmark leptoquark mass
of $M_{S_3}=3$~TeV (a value which satisfies the di-leptoquark production bound mentioned above).  
For this mass, we find the 95$\%$ CL limit from $pp\to \mu^+\mu^-$ to be weak,
essentially giving no further constraint on the best-fit region to flavour
data that we plot in Fig.~\ref{fig:param} (left panel). Specifically, for the
domain of $\theta_{23}$ values shown in that plot, the model is consistent with
$pp\to\mu^+\mu^-$ for $\lambda < 1.3$ or so ($\lambda=1.3$ is further to the
right of the plotted region). 

For the $\ZP$ model the present LHC constraints are also relevant.
Fig.~\ref{fig:param} 
displays 
95$\%$ CL lower limits upon $M_{Z^\prime}$
from CMS measurements of the di-muon mass
spectrum~\cite{CMS:2021ctt}. The
values of $g_{Z^\prime}/M_{Z^\prime}$ 
have been read off from
a re-casting of these measurements~\cite{Azatov:2022itm}, neglecting the 
dependence of the 
collider bounds upon $\theta_{23}$.
This is a good approximation because the dominant LHC signal process is $b
\bar b \rightarrow Z^\prime \rightarrow \mu^+\mu^-$, whose amplitude
is proportional to $\cos^2 \theta_{23} \approx 1 - \theta_{23}^2$ and
$-\theta_{23} < 0.18$ in the well-fit region. 

\subsection{Upper limits on the $\ZP$ and $S_3$ masses \label{sec:upper}}

Fig.~\ref{fig:param} shows that $\lambda \times 3 \mathrm{~TeV}/M_{S_3}
  > 0.03$ in the 95$\%$~CL region in the $S_3$ leptoquark model, which can be
  written $M_{S_3}/\text{TeV}<100\lambda$. 
  We then find an upper bound on $\lambda$ by considering
  the
  width-to-mass ratio of the $S_3$,
  which is
    $\Gamma_{S_3}/M_{S_3} = \lambda^2/(8\pi)$~\cite{Plehn:1997az,Dorsner:2016wpm}.
Requiring that the $S_3$ is perturbatively coupled, this width-to-mass ratio
should not be too large.
Requiring $\Gamma_{S_3}/M_{S_3} <1/3$ for perturbativity, we obtain $\lambda <
2.9$. Substituting this in, we find
$M_{S_3} < ~290\mathrm{~TeV}$,
from the combination of the fit to flavour data and perturbativity.
This is higher than previous estimates in
  Refs.~\cite{Allanach:2020kss,Azatov:2022itm} due in part to
  the LHCb December 2022 reanalysis of $R_{K^{(\ast)}}$, which made the
  minimum value on the abscissa smaller. The bound is also
  subject to large 
  fluctuations from the implementation of the $B_s -\overline{B_s}$
  mixing bound due to different lattice inputs for the SM prediction. 
Since the LHC di-leptoquark searches only constrain $M_{3}>1.4$~TeV, we see
that there is plenty of viable parameter space for a perturbatively coupled
$S_3$ that explains the \bsll\ anomalies.
The perturbativity and di-leptoquark searches bound combine to imply that
  $\lambda \times 3\text{~TeV}/M_{Z^\prime} \leq 6.2$ in the $S_3$ model.

Using a similar argument for the $\ZP$,
we can first infer from Fig.~\ref{fig:param}
that the 95\% CL region
  requires $g_{Z^\prime} \times 3 \mathrm{~TeV}/M_{\ZP} > 0.03$. 
The width of the $B_3-L_2$ $\ZP$ is approximately
$\Gamma_{\ZP}/M_{\ZP} = 13 g_{Z^\prime}^2/(8\pi)$~\cite{Allanach:2020kss}.
Taking $\Gamma_{\ZP}/M_{\ZP}< 1/3$, we find
$g_{Z^\prime}<0.80$, implying
the following upper bound on
the $\ZP$ mass coming from perturbativity and the global fit:
$M_{\ZP} < ~80\mathrm{~TeV}$,
higher than previous estimates in
Refs.~\cite{Allanach:2020kss,Azatov:2022itm}.
Given that the LHC constraints only exclude $M_{Z^\prime}$ up to
about 4.3~TeV, as indicated by the vertical purple dashed lines on the
right-hand plot 
of Fig.~\ref{fig:param}, we see that there is plenty of viable parameter
space for a perturbatively coupled $B_3-L_2$ $\ZP$ that explains the
\bsll\ anomalies.  
We see from Fig.~\ref{fig:param} and Ref.~\cite{Azatov:2022itm}, that in the region of 95$\%$ CL fit to
  flavour data, $M_{Z^\prime}
  \geq 2.0$~TeV. Combining this and perturbativity, we find that
  $g_{Z^\prime} \times 3\text{~TeV}/M_{Z^\prime} \leq 1.2$
  in the $Z^\prime$ model.

\section{Summary \label{sec:conc}}
We have contrasted two bottom-up beyond the SM physics models that can
significantly ameliorate the \bsll\ anomalies in global
fits\footnote{For a recent determination of the viable parameter spaces and future
collider sensitivities of the models, see Ref.~\cite{Azatov:2022itm}.}. 
Since one is a leptoquark model and one a $Z^\prime$ model, a naive
expectation is that the $Z^\prime$ model is disfavoured by the
fact that measurements of $B_s - {\overline B_s}$ mixing
are broadly compatible with the SM\@.
We
have shown that, 
contrary to the naive expectation, the $B_s - {\overline B_s}$ mixing
prediction of the best-fit point of each model
is similar,
being close to the SM prediction; $\Delta M_s$ is therefore not
a key discriminator.
That said, as Fig.~\ref{fig:param} shows, the $B_s-\overline{B_s}$ mixing constraint
  limits the $B_3-L_2$ model's value of $\theta_{23}$ in the $Z^\prime$
  model, albeit within a natural range of values that comfortably includes $\theta_{23} \approx |V_{cb}|$, whereas no such preference is strongly evident for the $S_3$ model.
The $Z^\prime$ model that we picked differs in other ways to the leptoquark
model in that it couples to di-muons through a vector-like coupling\footnote{We
note that it would be simple to change the $Z^\prime$ model such that the coupling
to di-muons is completely left-handed, by exchanging the $U(1)$ charges of $e_2
\leftrightarrow e_3$ in Table~\ref{tab:fields} in the fashion of
the Third Family Hypercharge Model~\cite{Allanach:2018lvl}.}, whereas
the leptoquark couples through a purely left-handed coupling. This shifts the
predictions for some of the other flavour observables around but in the end
the fits are of a similar quality to each other; the improvement in quality-of-fit
with respect to the SM of $\sqrt{\Delta \chi^2}=\Sthreedchisq{},
\BthreeLtwodchisq{}$ in the $S_3$ model and in the $Z^\prime$ model,
respectively.

The LHCb December 2022 reanalysis of $R_{K^{(\ast)}}$ pushes the fits to
  smaller new physics couplings to fermions for a given $b_L-s_L$ mixing
  angle. One unexpected consequence of including the reanalysed measurements
  was that the fit to each new physics model improves, as
  Table~\ref{tab:compare} displays. The improvement with respect to the SM
  decreases as expected, however each new physics model (with two fitted
  parameters) still improves upon the SM $\chi^2$ by some 12 units.
  This realisation should be tempered with the risk that the remaining
  discrepant $b \rightarrow s \mu^+ \mu^-$ observables could potentially
  receive unaccounted-for 
  long-distance contributions from charm penguins\footnote{\smelli{} ascribes 
  a large 
  theoretical uncertainty to $b\rightarrow s \mu^+\mu^-$ observable
  predictions in order to allow for this
  possibility.}. In
  Ref.~\cite{Ciuchini:2022wbq}, for example, the discrepancy with SM 
  predictions is ascribed completely to such contributions and is used to
  parameterise them.

As previously mentioned,
both in the \BthreeLtwo\ and in the $S_3$ leptoquark model, 
$B-$physics fits are insensitive to
scaling the coupling and the mass 
by the same
factor.
However, direct searches for the new hypothesised particles do not exhibit
this scaling symmetry. 
The $S_3$ model di-leptoquark search constraints
only depend sensitively on the mass of the leptoquark~\cite{Allanach:2019zfr},
since the production is governed by QCD and thus via the strong coupling
constant, which is of course known from experimental measurement.
Single leptoquark production in the $S_3$ model
does depend sensitively both upon the mass of the
leptoquark and upon its coupling~\cite{Allanach:2017bta}.
Direct $B_3-L_2$ $Z^\prime$ search constraints also depend upon the coupling
and the $Z^\prime$ mass~\cite{Allanach:2020kss}.
However, the scaling symmetry of the fits allows us in  \S\ref{sec:par}
to present complete fit constraints upon the three parameters of each model in a
two-dimensional plane. We hope that this will facilitate future direct
searches for either explanation of the \bsll\ anomalies.

\section*{Acknowledgements}
This work was partially supported by STFC HEP Consolidated grant
ST/T000694/1, by the SNF contract 200020-204428 and by the European Research
Council (ERC) under the European Union’s Horizon 2020 research and innovation
programme, grant agreement 833280 (FLAY). We thank Guy Wilkinson for prompting
this work with a seminar question in a LHCb UK seminar.
BCA thanks other members of the Cambridge Pheno Working Group for
helpful discussions and Peter Stangl for help with the development version of \smelli{}.
JD thanks Wolfgang Altmannshofer, Darius Faroughy, and Ben Stefanek for
discussions. We are very grateful to Admir Greljo for detailed feedback on the first version of this manuscript, and for drawing our attention to the complementary studies in Ref.~\cite{Azatov:2022itm}.

\bibliographystyle{JHEP-2}
\bibliography{rumble}

\end{document}